\documentclass[11pt]{article}
\pdfoutput=1
\usepackage{jheppub}
\usepackage{upgreek}
\usepackage{graphicx}
\usepackage{slashed}
\usepackage{amssymb}
\usepackage{subcaption}
\usepackage{amsmath,amssymb}
\usepackage{slashed}
\usepackage{xcolor}
\usepackage{dsfont}
\usepackage{verbatim}
\usepackage{mathtools,xcolor,ytableau,amsfonts,tikz}
\usepackage{physics}
\usepackage{simpler-wick}
\usepackage[nobottomtitles*]{titlesec}
\usepackage{color,soul}
\usepackage{cancel}
\setulcolor{orange}

\numberwithin{equation}{section}

\newcommand{\mL}{\mathcal{L}}
\newcommand{\mO}{\mathcal{O}}
\newcommand{\pd}{\partial}

\newcommand\beq{\begin{equation}}
\newcommand\eeq{\end{equation}}

\author[a,b]{Gabriel Cuomo,}
\author[c]{Leonardo Rastelli,}
\author[d]{Adar Sharon}

\affiliation[a]{Center for Cosmology and Particle Physics, Department of Physics, New York University, New York, NY 10003, USA}
\affiliation[b]{Department of Physics, Princeton University, Princeton, NJ  08544, USA}
\affiliation[c]{C. N. Yang Institute for Theoretical Physics, Stony Brook University, Stony Brook, NY 11794, USA}
\affiliation[d]{Simons Center for Geometry and Physics, SUNY, Stony Brook, NY 11794, USA}

\emailAdd{gc6696@princeton.edu}			 
\emailAdd{leonardo.rastelli@stonybrook.edu}
\emailAdd{asharon@scgp.stonybrook.edu}    

\title{\LARGE{Moduli Spaces  in CFT:}\\ \LARGE{Bootstrap Equation in a Perturbative Example}}

\abstract{
Conformal field theories that exhibit 
spontaneous breaking of conformal symmetry 
(a moduli space of vacua) must satisfy
a set of bootstrap constraints,  involving the usual data (scaling dimensions and OPE coefficients) as well as new data such as the spectrum of asymptotic states in the broken vacuum and form factors. 
The simplest bootstrap equation arises by 
expanding a two-point function of local operators in two channels,  at short distance using the OPE and at large distance using the EFT in the broken vacuum. We illustrate this equation in what is arguably the simplest perturbative model that exhibits conformal 
symmetry breaking, namely the real $ABC$ model
in $d = 4 -\epsilon$ dimensions. We investigate the convergence properties of the bootstrap equation and check explicitly many of the non-trivial relations that it imposes on  theory data. 
}

\begin{document}

\maketitle

\section{Introduction}

Spontaneous breaking of continuous global symmetries  is a ubiquitous phenomenon  in quantum field theory. As is well known, it can occur in $d >2$ spacetime dimensions, where it leads to massless Goldstone bosons that parametrize the flat directions of the moduli space of vacua. By contrast, spontaneous breaking of {\it conformal} symmetry appears to be very non-generic. All known interacting examples\footnote{
We restrict our analysis  to legitimate local CFTs, in particular we require the existence of a local stress-tensor operator with finite two-point function. It is not difficult to engineer examples of flat directions in (non-supersymmetric) CFTs to leading order in the large $N$ expansion, see~e.g.~\cite{Rabinovici:1987tf,PhysRevLett.52.1188,Kondo:1988qd,Chai:2020zgq,Chai:2020onq,Cresswell-Hogg:2022lez,Semenoff:2024prf} or (which is much the same) in theories defined by a classical AdS dual \cite{Bellazzini:2013fga,Coradeschi:2013gda,Faedo:2024zib}. These moduli spaces are expected to get lifted at finite $N$. 
} are supersymmetric,
but it is not  clear whether this is a deep structural fact or just a lamppost effect.
Intuitively, conformal symmetry breaking is rare (and perhaps always requires supersymmetry) because  a delicate conspiracy is needed to fine tune to zero the potential of the dilaton (the Goldstone boson of broken scale invariance) \cite{Salam:1969bwb,Isham:1970gz}.
It is however difficult to phrase this intuition in abstract CFT language.
 Ideally, one would like a simple sharp criterion that establishes which CFTs (abstractly defined by their usual conformal bootstrap data) admit moduli spaces of vacua. 

How to formulate the abstract bootstrap question is conceptually clear, if not immediately useful. 
CFTs that admit a moduli space of vacua have a larger set of observables, which must satisfy some stringent bootstrap conditions. One can draw an illuminating analogy with  CFTs in the presence of a conformal boundary. 
In boundary CFT, there are three sets of data: 
\begin{enumerate}
\item[(i)] The usual bulk data: spectrum and OPE coefficients of the bulk local  operators.
\item[(ii)] Boundary data: spectrum and OPE coefficients of the boundary local operators.
\item[(iii)] Bulk-to-boundary data. Using the OPE, they can be reduced to the two-point functions of one bulk and one boundary operator.
\end{enumerate}
In a CFT with a moduli space of vacua,  there are also three sets of data:
\begin{enumerate}
\item[(i)'] The usual ``bulk'' data: spectrum and OPE coefficients of the ``bulk'' local operators.
\item[(ii)'] S-matrix bootstrap data: the spectrum of asymptotic states in the broken vacuum, and their complete S-matrix. The massless dilaton is always one of the asymptotic states, possibly accompanied by additional massless moduli. 
\item[(iii)'] Form factors, i.e.~the overlaps between the local operators and  the asymptotic states in the broken vacuum.
\end{enumerate}
In both cases, one is breaking the original $SO(d, 2)$ conformal symmetry to a subgroup, which is $SO(d-1, 2)$
in BCFT and the  Poincaré group $ISO(d-1, 1)$ for a CFT on its moduli space. In both cases, the original CFT data are supplemented by  new sets of data, which 
are specific to the  choice of boundary state  or to the choice of broken vacuum, respectively. The new data
must obey stringent bootstrap constraints. 
But the analogy only goes so far. First, there are profound technical differences, as BCFT requires only a relatively minor extension of the conformal bootstrap framework, while CFTs on their moduli space present us with a novel interplay between the conformal bootstrap  and the S-matrix bootstrap. Second,  there is an essential physical difference: a generic CFT is expected to admit at least one  consistent boundary state,\footnote{More precisely, the general lore is that all CFTs {\it with no gravitational anomaly} can be put on a space with boundary.} but {\it no} moduli space of vacua. The  problem of conformal symmetry breaking can then be precisely phrased: for which special choices of CFT data (i)'
 can one find  data (ii)' and (iii)' that satisfy all the bootstrap constraints?

In this paper we will not be able to answer this difficult question. Our more modest aim will be to illustrate the simplest bootstrap equation,  namely the one
that arises from the two-point function of scalar primary operators~\cite{Karananas:2017zrg}, in a very concrete example. In a companion paper~\cite{Cuomo:2024fuy}, we address the question from a very different angle, providing a simple {\it necessary} condition for spontaneous conformal symmetry breaking, under the assumption that a continuous global symmetry is {\it also} broken  on the moduli space.
 
As we have emphasized, all known interacting CFTs with a moduli space are supersymmetric. The most familiar examples are theories with at least four real supercharges in $d=3$ and $d=4$ spacetime dimensions. Their moduli spaces are holomorphic, parametrized by expectation values of chiral operators which are charged under a continuous R-symmetry.  
Moduli spaces can also occur in 3d SCFTs with only two real supercharges, as a $\mathbb{Z}_2$ R-symmetry is sometimes sufficient to establish the existence of flat directions (see \cite{Gaiotto:2018yjh} for a recent discussion).
These models are particularly interesting for our purposes. As the powerful restrictions imposed  by holomorphy are absent, their low-energy EFTs are less constrained, and their operator spectrum more generic -- in  particular there are no towers of protected chiral operators.  We may thus hope to learn more general lessons. A related benefit is the simplicity of several Lagrangian models in this class. We focus on what is arguably the simplest perturbative example that exhibits conformal symmetry breaking, namely the real $ABC$ model in $d=4 - \epsilon$ dimensions. Its field content is comprised of three real scalars and three real fermions.\footnote{This is the counting of degrees of freedom in the physical dimension $d=3$. To perform the $\epsilon$-expansion one needs to analytically continue in the number of fermions~\cite{Fei:2016sgs,ThomasSeminar}, as we will review in detail.} The moduli space consists of three equivalent branches (exchanged by the discrete global symmetry)  that meet at the origin, each branch corresponding to one of the three scalars
 acquiring a vacuum expectation value. The real $ABC$ model illustrates conformal symmetry breaking in the minimal setting where the dilaton is the only massless modulus.

A basic set of consistency conditions for a CFT with a moduli space arises by considering a two-point function of local operators 
and expanding it in 
 two ``channels'',
 \begin{eqnarray} \label{bootstrap}
\langle 0 | {\cal O}_i (x) {\cal O}_j (0) |0 \rangle  & = &
\sum_k \frac{g_{ij}^{\;k}  }{  |x|^{\Delta_i + \Delta_j - \Delta_k}}\; \langle 0| {\cal O}_k (0) | 0\rangle \qquad {\rm for}\; x \to 0\\
& = & \sum_n  \langle 0 | {\cal O}_i (x) |n \rangle \langle n | {\cal O}_j (0) |0 \rangle   
\, \quad \qquad {\rm for}\; x \to \infty \,. \nonumber 
     \end{eqnarray}
     At short distance, we use the OPE and take the vacuum expectation value term by term.
 While in the standard conformal vacuum only the identity term contributes, in the broken vacuum scalar primaries can acquire a vev. The large distance expansion is instead obtained by  inserting the complete set of asymptotic states $\{ |n \rangle \}$ in the broken vacuum.  This equation was first written down in~\cite{Karananas:2017zrg} but to the best of our knowledge never studied in any detail. 
 
 A first structural question is 
 about convergence of the two expansions. 
It seems plausible that the short distance expansion is absolutely convergence for any $|x| < \infty$. We believe this to be the case at least in all perturbative models. Indeed the short distance expansion in the broken vacuum is just the usual OPE, with the  insertion at $x = \infty$ of a sum of scalar primaries that enforce the non-zero vevs. A sufficient condition for its convergence is that the state at infinity has finite norm; this is true in free field theory and we will argue that it remains true to all orders in perturbation theory. By contrast,  the $x \to \infty$ expansion is only asymptotic, as can be checked already at tree level in our simple example. In momentum space, both the short and the large distance expansions might have 
finite radius of convergence, but there is no overlapping region where they both converge. This complicates the use of the equation in an abstract bootstrap setting, but it is not an obstacle in our concrete example, where we can just fully compute both sides of \eqref{bootstrap} to any  given order in perturbation theory.

We compute the two-point functions of the elementary fields up to one loop order in the real $ABC$ model, in a double-scaling limit where the mass of the single-particle states  is kept fixed.
We find that that \eqref{bootstrap} implies intricate constraints on the data of the theory. Already at tree level, where the large distance expansion
is very simple (only single particle states contribute, either the massless dilaton or a massive scalar), matching with the short distance expansion is non-obvious -- even the basic requirement that there is no divergence as $x \to \infty$  requires large cancellations.
While these relations amongst theory data ultimately follow from the selection rules enforced by supersymmetry on the Lagrangian model, they are definitely not a consequence of superconformal representation theory alone, and  appear highly non-trivial from an abstract bootstrap viewpoint.

\medskip
\noindent
 The remainder of the paper is organized as follows. In section~\ref{sec_crossing}
 we introduce the basic bootstrap equation \eqref{bootstrap} and investigate its convergence properties. We also prove that infinitely many scalar primaries must acquire a vev. In section~\ref{sec_ABC}
 we briefly review 3d ${\cal N}=1$ models, 
  their moduli spaces, and how to set-up their $\epsilon$-expansion, focusing on the example of the real $ABC$ model.
Section~\ref{sec_bootABC} is the main technical part of the paper. We compute two-point functions of elementary fields in the 
 real $ABC$ model up to one loop order.  We illustrate how the bootstrap equation \eqref{bootstrap} works in this concrete example and check many of the non-trivial relations that it implies on the data of theory.
We conclude in section~\ref{sec_outlook} with a brief outlook.
Appendix~\ref{app_ABC_CFT_data} contains several technical details of our perturbative calculations. 

\medskip
\noindent
\textbf{Note added:} As we were finalizing this work, \cite{Ivanovskiy:2024vel} appeared on the ArXiv. The authors study similar questions in the context of four dimensional planar $\mathcal{N}=4$ Super-Yang-Mills theory. Our results for the convergence of the OPE agree with the explicit calculation there.

\section{Bootstrap equation}\label{sec_crossing}

In this section we review and study the basic bootstrap equation (\ref{bootstrap}). Our main new contribution is a discussion of its convergence properties.

\subsection{OPE on the moduli space}\label{subsec_crossing_intro}

Let us first introduce some notation. We consider a generic CFT in $d>2$. We denote its (possibly trivial) internal symmetry group $G$. We suppose that the theory admits an $n$-dimensional moduli space in which infinitely many\footnote{As we will discuss shortly, consistency of the OPE implies that, once a scalar operator with dimension $\Delta>0$ admits a nontrivial one-point function, infinitely many other operators also acquire a vev.} scalar primary operators acquire a vev and break the conformal symmetry, as well as the internal symmetry as $G\rightarrow H$. We assume that the moduli space is $n$-dimensional, i.e.~that the vacuum we are considering can be specified in terms of the expectation values $\langle\mO_i\rangle=v_i$ of $n$ operators, which thus provide a set of local coordinates on the moduli space. Notice that the dimension of the moduli space is always larger or equal to $n_{G/H}+1$, where $n_{G/H}$ is the dimension of the coset $G/H$. At a generic point on the moduli space 
it is always possible to make a change of coordinates $\{v_i\}\rightarrow \{v,\phi_A\}$ with $A=1,\ldots, n-1$, such that the $\phi_A$ take values on a compact manifold while $v>0$ is noncompact. Therefore, choosing $v$ to have mass dimension one, we write the expectation value of an arbitrary real scalar primary operator as
\begin{equation}\label{eq_vacua}
\langle  \mO_{\Delta,X}(x)\rangle_{v,\vec{\phi}}\equiv \langle v,\vec{\phi}| \mO_{\Delta,X}(x)|v,\vec{\phi}\rangle=\xi_{\mO} v^\Delta\,.
\end{equation}
In~\eqref{eq_vacua}, $(\Delta,X)$ are the scaling dimension and the other internal quantum numbers of the scalar primary operator, and the $\{\xi_{\mO}\}$ are a set of nontrivial coefficients that may depend on the exactly marginal couplings $\sim g$ of the theory (the so-called conformal manifold) and on the parameters of the moduli space: $\xi_{\mO}=\xi_{\mO}(\vec{\phi},g)$. We will drop the subscript $v,\vec{\phi}$ in what follows. 

In CFTs, local operators form a closed algebra under the operator product expansion (OPE). As is well known, this allows to formally reduce all correlation functions to a sum of one-point functions by repeated fusion of the operators. The only difference between the conformal vacuum and the broken vacuum is that, in the first, only the identity operator admits a nontrivial one-point function, while in the latter all scalar primary operators may acquire a nontrivial vev. In this section we explore some of the consequence of the OPE for the CFT data under the assumption that the theory admits a moduli space.

Let us warm up with two simple general considerations.
First, the vacua which exhibit spontaneous symmetry breaking (SSB) can obviously be seen as a consequence of different, nontrivial, boundary conditions at $x\rightarrow\infty$. In radial quantization we can thus regard correlation functions on the moduli spaces as matrix elements between the conformal vacuum at $|x|=0$ and a nontrivial state ${}_{\rm rad}\langle v,\vec{\phi} |$ at $|x|=\infty$. The OPE allows us to formally deduce such state from the requirement that all one-point functions are correctly reproduced. We find
\begin{equation}\label{eq_Ishibashi}
{}_{\rm rad}\langle v,\vec{\phi} |={}_{\rm rad}\langle 0|\sum_{\Delta,X} \mO_{\Delta,X}(\infty)\langle\mO_{\Delta,X}\rangle_{v,\vec{\phi}}\,,
\end{equation}
where as usual $\mO(\infty)=\lim_{x\rightarrow\infty}|x|^{2\Delta}\mO(x)$ and we assume canonically normalized operators $\langle 0|\mO_i(\infty)\mO_j(x)|0\rangle=\delta_{ij}$.
Equation \eqref{eq_Ishibashi} is formally analogous to the structure of Cardy states in boundary CFTs~\cite{Cardy:2004hm}.

As we will see, upon using the OPE most correlation functions reduce to a sum of infinite terms. In some cases however the OPE truncates. Most notably, this is the case for the two-point functions of the stress tensor $T^{\mu\nu}$ or a conserved current $j^\mu_a$ with a scalar primary operator $\mO_{\Delta,X}(0)$. This is because  the OPE takes the form
\begin{align}\label{eq_OPE_jO}
&j^\mu_a(x)\mO_{\Delta,X}(0)\sim \frac{x^\mu}{\Omega_{d-1}|x|^d}\delta_a\mO_{\Delta,X}(0)+\text{descendants} +\text{spinning operators}\,,\\
& T^{\mu\nu}(x)\mO_{\Delta,X}(0)\sim\frac{d\Delta}{d-1}\frac{\delta^{\mu\nu}-\frac{x^\mu x^\nu}{d x^2}}{\Omega_{d-1}|x|^d}\mO_{\Delta,X}(0)
+\text{descendants} +\text{spinning operators}
 \,,
 \label{eq_OPE_TO}
\end{align}
where $\Omega_{d-1}=\frac{2 \pi ^{d/2}}{\Gamma \left(d/2\right)}$ is the volume of the $d-1$-dimensional sphere, and $\delta_a\mO_{\Delta,X}=[Q_a,\mO_{\Delta,X}]$ is the variation of the operator under the internal generator $Q_a$ associated with $j_a^\mu$, e.g. $\delta\mO_{\Delta,q}=q\mO_{\Delta,q}$ for an internal $U(1)$ symmetry and a charge $q$ operator. The absence of additional scalar primaries in \eqref{eq_OPE_jO} and~\eqref{eq_OPE_TO} follows from the Ward identities, which imply $\pd_\mu j^\mu_a(x)\mO_{\Delta,X}(0)=\delta^d(x)\delta_a\mO_{\Delta,X}(0)$ and $T^\mu_{\,\mu}(x)\mO_{\Delta,X}(0)=\delta^d(x)\Delta\mO_{\Delta,X}(0)$.
Since descendants and spinning operators have vanishing expectation value, the OPE relations~\eqref{eq_OPE_jO} and~\eqref{eq_OPE_TO} imply that the two-point functions of the current and the energy-momentum tensor with the order parameter read
\begin{align}\label{eq_jO2pt}
&\langle j_a^\mu(x)\mO_{\Delta,X}(0)\rangle=\frac{x^\mu}{\Omega_{d-1}|x|^d}
\langle\delta_a\mO_{\Delta,X}\rangle=
\frac{x^\mu}{\Omega_{d-1}|x|^d} (Q_a\xi_\mO) v^{\Delta}\,,\\
&\langle T^{\mu\nu}(x)\mO_{\Delta,X}(0)\rangle=
\frac{d\Delta}{d-1}\frac{\delta^{\mu\nu}-\frac{x^\mu x^\nu}{d x^2}}{\Omega_{d-1}|x|^d}\langle \mO_{\Delta,X}\rangle
=\frac{d\Delta}{d-1}\xi_{\mO}v^\Delta\frac{\delta^{\mu\nu}-\frac{x^\mu x^\nu}{d x^2}}{\Omega_{d-1}|x|^d}\,.
\label{eq_TO2pt}
\end{align}
As is well known, transforming to momentum space \eqref{eq_jO2pt} and~\eqref{eq_TO2pt} one finds that both correlators display a pole $\sim 1/p^2$.  This proves proves the existence of $1+n_{G/H}$ massless particles, the dilaton and the Goldstone Bosons for the broken internal symmetry.

Let us now focus on the main object of interest in this paper: two-point functions of primary operators
\begin{equation}\label{eq_2pt_OO}
\langle\mO_i(x)\mO_j(0)\rangle\,,
\end{equation}
where we collectively denoted with subscripts $i,j$ the relevant quantum numbers. For simplicity, we restrict to {\it scalar} primaries. We also choose  a basis where all operators are real: $\mO_i=\mO_i^\dagger$.

The correlation function~\eqref{eq_2pt_OO} satisfies a bootstrap equation, first presented in \cite{Karananas:2017zrg}. On the one hand, \eqref{eq_2pt_OO} can be evaluated using the OPE  
\begin{equation}\label{eq_OO_OPE}
\mO_{i}(x)\mO_{j}(0)=\frac{\mathcal{N}_{ij}}{|x|^{2\Delta}}+\sum_{k} g_{ij}^{\:\:\: k}
\frac{\mO_k(0)}{|x|^{\Delta_i+\Delta_j-\Delta_k}}+\text{descendants}+\text{spinning operators}\,,
\end{equation}
where the first term is the contribution of the identity operator\footnote{For generality, from now on we do not assume that the operators are canonically normalized.} and $g_{ij}^{\:\:k}$ denotes the OPE coefficients (which are constrained by the internal symmetry). On the other hand,  we can write~\eqref{eq_2pt_OO} (at finite distance) as the sum of the disconnected term and the K\"allen-Lehman decomposition of the connected correlator
\begin{equation}\label{eq_KL_dec}
\langle\mO_i(x)\mO_j(0)\rangle=\xi_i\xi_jv^{\Delta_i+\Delta_j}+
\int dm^2\rho_{ij}(m^2) G_{m^2}(x)\,,
\end{equation}
where we used $\langle \mO_i\rangle=\xi_i v^{\Delta_i}$, $G_{m^2}(x)$ is the free propagator
\begin{equation}\label{eq_scalar_prop}
G_{m^2}(x)=
\int \frac{d^dp}{(2\pi)^d}\frac{e^{ipx}}{p^2+m^2}=
\frac{m^{d-2}}{(2\pi)^{d/2}(m|x|)^{\frac{d-2}{2}}}K_{\frac{d-2}{2}}(m|x|)
\,,
\end{equation}
and $\rho_{ij}$ is the spectral density 
\begin{equation}\label{eq_spectral_density}
\rho_{ij}(p^2)=(2\pi)^{d-1}\sum_{n}\langle 0|\mO_i|n\rangle\langle n|\mO_j|0\rangle\delta^d(q_n-p)\,.
\end{equation}
Combining~\eqref{eq_OO_OPE} and~\eqref{eq_KL_dec} we obtain the relation
\begin{equation}\label{eq_master}
\sum_{k} g_{ij}^{\:\:\: k}\xi_k L^{\Delta_k-\Delta_i-\Delta_j}-\xi_i\xi_j=v^{-\Delta_i-\Delta_j}\int dm^2\rho_{ij}(m^2) G_{m^2}(x)\,,\quad
L=v|x|\,.
\end{equation} 
This equation relates the mass spectrum on the moduli space to the CFT data in the conformal vacuum.\footnote{Analogous equations can be written in other setups, e.g.~for thermal correlators~\cite{Caron-Huot:2009ypo,El-Showk:2011yvt} or correlation functions in a finite density sate. In these other examples, however, the spectral density decomposition is not constrained by relativity and spinning operators may also acquire a vev $\langle \mO_{\mu_1\ldots\mu_n}\rangle\propto\delta_{\mu_1}^0\ldots\delta_{\mu_n}^0$.} We will refer to it as the bootstrap or crossing equation.

As a first trivial application,  we use~\eqref{eq_master} to prove that infinitely many operators must acquire a vev in the broken vacuum. If the sum on the left-hand side is infinite there is nothing to prove. If instead it truncates, \eqref{eq_master} implies that there must be at least one operator with scaling dimension $\Delta_k=\Delta_i+\Delta_j$. This is because the massless particles on the moduli space together with the cluster decomposition principle and unitarity imply that, at large distances, the connected two-point function behaves as
\begin{equation}\label{eq_def_Pi}
\langle\mO_i(x)\mO_j(0)\rangle_c=\int dm^2\rho_{ij}(m^2) G_{m^2}(x)\stackrel{x\rightarrow\infty}{\sim}\Pi_{ij}\frac{v^{\Delta_i+\Delta_j-(d-2)}}{|x|^{d-2}}+\ldots
\,,
\end{equation}
where $\Pi_{ij}$ is a (potentially vanishing) coefficient that depends upon the matrix element of the operators with the massless scalar particles $\phi_a$ of the moduli space and the dots stand for terms which vanish faster than $1/|x|^{d-2}$ for large $x$.   
Therefore in the limit $L\rightarrow\infty$,~\eqref{eq_master} reads
\begin{equation}\label{eq_master_long}
\lim_{L\rightarrow\infty}\sum_{k} g_{ij}^{\:\:\: k}\xi_k L^{\Delta_k-\Delta_i-\Delta_j}=\xi_i\xi_j\,,
\end{equation}
which, since the sum truncates by assumption,  implies that all operators satisfy $\Delta_k\leq \Delta_i+\Delta_j$, and that there there is at least an operator with scaling dimension $\Delta_k=\Delta_i+\Delta_j$. We may then iterate the argument for the two-point function of such operator, and so on, and conclude that infinitely many operators, with arbitrary large scaling dimension, must acquire a vev.

In some cases, the sum on the left-hand side of~\eqref{eq_master} does truncate for a family of correlators. This is famously the case in free theories or for holomorphic correlation functions of chiral operators in SCFTs. In this case the previous argument implies the existence of a ring of operators with scaling dimension $\sum_k n_k\Delta_k$, where $n_k\in\mathds{N}$ and  $\Delta_k$ are the dimensions of the generators of the ring.

As we will see, in most cases the sum in~\eqref{eq_master} receives contributions from an infinite number of terms.  In this case we cannot commute the sum with the limit in~\eqref{eq_master_long}, and there is no need for an operator with scaling dimension $\Delta_i+\Delta_j$ to exist. 

\subsection{Short and long distance expansions}\label{subsec_convergence}

To analyze the consequences of~\eqref{eq_master} we should understand the convergence properties of both the short and long distance expansions on the moduli space.  Here we provide some general comments. We will verify our statements in a concrete example in the next sections.

Let us first discuss the convergence of the short distance expansion.
To this aim, we can use the Cardy-like representation in~\eqref{eq_Ishibashi} to write the two-point functions in radial quantization as
\begin{equation}\label{eq_2pt_radial}
\begin{split}
\langle\mO_i(x)\mO_j(0)\rangle &=
{}_{\rm rad}\langle v,\vec{\phi}|\mO_i(x)\mO_j(0)|0\rangle_{\rm rad}\\
&=\sum_{l,k}\left(\mathcal{N}^{-1}\right)_{lk}\xi_k v^{\Delta_k}{}_{\rm rad}\langle 0| \mO_{k}(\infty)\mO_i(x)\mO_j(0)|0\rangle_{\rm rad}\,.
\end{split}
\end{equation}
This equation suggests that the OPE converges for any $|x|<\infty$ by the usual argument \cite{Rychkov:2016iqz}. Indeed the product $\mO_i(x)\mO_j(0)|0\rangle_{\rm rad}$ corresponds to a state in radial quantization. By the state operator correspondence this state can  be written as a linear combination of primaries and descendants. 

To make the former argument argument rigorous a sufficient (but actually not necessary) condition is that the state ${}_{\rm rad}\langle v,\vec{\phi}|$ is normalizable. As an example, it is simple check to that this is the case in the theory of a free scalar $\phi$ in $d$-dimensions. Indeed at leading order the only operators acquiring a vev are those of the form $\phi^n$. Since their tree-level two-point function in the conformal vacuum is
\begin{equation}
    \langle \phi^n(x)\phi^n(0)\rangle_{v=0}=n! [G_{0}(x)]^n=\frac{n!}{(d-2)^n\Omega_{d-1}^n|x|^{2n}}\,,
\end{equation}
the corresponding normalizable state is $|\hat{\phi}^n\rangle=\sqrt{\frac{[G_0(1)]^n}{n!}}\phi^n|0\rangle=\frac{\sqrt{(d-2)^n\Omega_{d-1}^n}}{\sqrt{n!}}\phi^n|0\rangle$. Therefore we find that the Cardy-like state~\eqref{eq_Ishibashi} reads
\begin{equation}
 {}_{\rm rad}\langle v|=\sum_{n=0}^{\infty}v^{n\frac{d-2}{2}}\sqrt{\frac{[G_{0}(1)]^n}{n!}} \langle \hat{\phi}^n|\quad  \implies\quad
 |v\rangle_{\rm rad}=
 \sum_{n=0}^{\infty}R^{n(d-2)}v^{n\frac{d-2}{2}}
 \sqrt{\frac{[G_{0}(1)]^n}{n!}}| \hat{\phi}^n\rangle\,,
\end{equation}
where $v^{\frac{d-2}{2}}=\langle\phi\rangle$ and in the formula for the conjugate state $R$ is an arbitrary distance scale, which can be thought as the distance between the origin and the point around which we perform the inversion in the definition of a conjugate operator in radial quantization. We conclude that the norm of the state is finite
\begin{equation}\label{eq_Ishibashi_norm}
    {}_{\rm rad}\langle v|v\rangle_{\rm rad}=\sum_{n=0}^{\infty}\frac{\left[G_{0}(1)(R v)^{d-2}\right]^n}{n!}=\exp\left[\frac{(R v)^{d-2}}{(d-2)\Omega_{d-1}}\right]\,.
\end{equation}
In perturbative theories, such as $\mathcal{N}=4$ SYM at weak coupling, higher orders in perturbation theory are not expected to affect the finiteness of the norm~\eqref{eq_Ishibashi_norm}, and thus the convergence of the OPE. Indeed, the $k$-th loop correction to $\langle \phi^n\rangle$ scales at most as $g^{2k}n^{2k}$, where $g$ is a cubic coupling (the estimate is reliable for $1\ll k\ll n$). Thus loops do not modify the $\sim 1/\sqrt{n!}$ behaviour of the expectation value of the normalized operator for large $n$; similarly, the norm of operators with many derivatives is also suppressed by large factorials.  Given these arguments, we expect that the OPE in the moduli space converges at any finite $|x|$ to all orders in perturbation theory, and more in general in all models that admit a perturbative limit.

Let us now discuss the long distance limit. As we observed earlier around~\eqref{eq_def_Pi}, this limit is controlled by the cluster decomposition principle and the massless particles of the moduli space. In general, the long distance expansion contains both inverse powers of $|x|$, which are generically computed from the EFT for the massless fields in the moduli space, as well as exponential corrections due to the exchange of massive modes. Therefore the long distance expansion is always asymptotic.  The requirement that correlation functions remain finite at large separation implies that the series that we obtain using the OPE must have oscillating signs. This is because a  series with positive coefficients that converges for any $|x|<\infty$ necessarily gives rise to a function that diverges for $|x|\rightarrow\infty$.

These abstract considerations are concretely exemplified by the exchange of a single mass $m$ particle in the two-point function.\footnote{In many perturbative examples, such as in $\mathcal{N}=4$ and the $ABC$ model that we discuss in the rest of the paper,  correlation functions at leading order in the coupling are given by a finite sum of free massless and massive propagators.} This results into a contribution proportional to the free propagator
\begin{equation}
\langle \mO(x)\mO(0)\rangle\supset \#
\frac{m^{d-2}}{(2\pi)^{d/2}(m|x|)^{\frac{d-2}{2}}}K_{\frac{d-2}{2}}(m|x|)\,.
\end{equation}
It follows from the properties of the Bessel function that the short distance expansion of the massive propagator for $m|x|\ll1 $ is uniformly convergent for any finite $x^2$. Additionally, it is well known that the expansion of the Bessel function $K_a(z)$ is given by an oscillating series. E.g. in $d=4$ we have
\begin{equation}\label{eq_ABC_Bessel_series}
K_1(z)=\frac{1}{z}+\sum_{k=0}^\infty \left(\frac{z}{2}\right)^{2k+1}\left[p_k\log\left(\frac{z}{2}\right)+q_k\right]\,,
\end{equation}
where the $p_k$'s and $q_k$'s have opposite sign
\begin{equation}\label{eq_ABC_Bessel2}
p_k=\frac{1}{k! (k+1)!}\,,\qquad
q_k=-\frac{\psi ^{(0)}(k+1)+\psi ^{(0)}(k+2)}{2\left[k! (k+1)!\right]}\stackrel{k\rightarrow\infty}{\sim}-\frac{\log k}{k! (k+1)!}\,.
\end{equation}
As we will see, the logarithms in~\eqref{eq_ABC_Bessel_series} are interpreted as perturbative contributions from the anomalous dimensions of the operators exchanged in the OPE in concrete models.
Both the sum over the $p_k$'s and the $q_k$'s produce,  individually, a behaviour of the sort $\sim e^{z}\log(z)/z$ for large $z$. However such contributions cancel when summing them up and the end-result decays exponentially $\sim e^{-z}/z$.

Incidentally, from~\eqref{eq_ABC_Bessel2} we see that, in order to reproduce the exchange of a finite number of massive particles in $d=4$, we must have operators with scaling dimension $\Delta\simeq 2n-2$ with $n\in\mathds{N}$, whose product of OPE coefficients and one-point functions produces two terms scaling as $\sum\xi_{\mO_{\Delta}} g_{\mO\mO}^{\;\;\mO_{\Delta}}\sim 1/[\Gamma(\Delta/2)]^2\sim e^{-\Delta\log\Delta} $.  More generally, a simple saddle-point argument suggests that that the regularity of the lond distance limit implies that the sum over operators of dimension close $\Delta$ yields $\sum\xi_{\mO_{\Delta}} g_{\mO\mO}^{\;\;\mO_{\Delta}}\sim e^{-i\pi \Delta-\Delta \log\Delta}$ \cite{Ivanovskiy:2024vel}. This scaling is not generic and implies nontrivial cancellations between the CFT data of the theory. Indeed, from the discussion above we infer that one-point functions of scalar operators scale as $\sim 1/\sqrt{\Gamma(\Delta)}$ for large $\Delta$ in perturbative models, while from \cite{Bhattacharyya:2007vs,Shaghoulian:2015lcn,Mukhametzhanov:2018zja,Benjamin:2023qsc} we know that in CFT the product of the density $\rho(\Delta)$ of scalar primary operators $\mO_{\Delta}$ and OPE coefficients $g_{\mO\mO}^{\;\;\mO_{\Delta}}$ scale as $\rho(\Delta)|g_{\mO\mO}^{\;\;\mO_{\Delta}}|\sim 2^{-\Delta}$. This would suggest a scaling $\sum\xi_{\mO_{\Delta}} g_{\mO\mO}^{\;\;\mO_{\Delta}}\sim 1/\sqrt{\Gamma(\Delta)}$ for the coefficients in~\eqref{eq_ABC_Bessel2}, which is much larger than the actual result.

Finally, it is also possible to consider two-point functions in momentum space.  We do not have general arguments for the convergence of the small and large momentum expansion. The example of a free massive propagator $(p^2+m^2)^{-1}$ suggests that these expansions in general have at most a measure zero overlapping regime of convergence. We will see that this conclusion holds also at one-loop level in the example that we will study in section~\ref{sec_bootABC}.

\section{Moduli spaces in \texorpdfstring{$3d$ $\mathcal{N}=1$}{3d N=1} SCFTs and the real \texorpdfstring{$ABC$}{ABC} model}\label{sec_ABC}

In this section we first review some general facts about three-dimensional ${\cal N}=1$ theories and their $\epsilon$-expansion, 
and then focus on the real $ABC$ model, arguably the simplest perturbative Lagrangian model that admits conformal symmetry breaking.

\subsection{Moduli spaces in \texorpdfstring{ $3d$ $\mathcal{N}=1$}{3d N=1} theories}\label{subsec_ex_3d_SUSY}

Three-dimensional theories with $\mathcal{N}=1$ SUSY do not have a continuous R-symmetry, and so lack the full power of SUSY that theories with more than two supercharges possess. However, it turns out that a discrete R-symmetry is enough in order to provide some exact results, and in particular in some cases it is enough to protect a moduli space from being lifted due to quantum corrections. Therefore, $3d$ $\mathcal{N}=1$ SCFTs are the minimal known interacting theories that admit moduli spaces, since all one needs are two supercharges and some discrete internal symmetry group. These ideas have appeared in various places in the literature, e.g.~\cite{Affleck:1982as,Gremm:1999su,Gukov:2002es}, and we review them mostly following the recent discussion in \cite{Gaiotto:2018yjh}. 

Our $3d$ $\mathcal{N}=1$ superspace conventions follow \cite{Gates:1983nr}. A Lagrangian for a $3d$ $\mathcal{N}=1$ SUSY theory can be written in superspace, with $x^\mu$ the usual coordinates and $\theta_\alpha$ Majorana Grassmannian coordinates. The superpotential takes the form
\begin{equation}
    \int d^2\theta W(\Phi,\bar{\Phi})\;,
\end{equation}
where we have emphasized that $W$ is not a holomorphic function of the fields and that it does not include derivatives. 
The basic observation of \cite{Affleck:1982as,Gaiotto:2018yjh} is that the superspace coordinates behave as Majorana fields, and therefore transform as $d^2\theta\to -d^2\theta$ under time-reversal $T$. For the theory to be  invariant under $T$, the superpotential must be odd under it.\footnote{In components, this requirement follows from the transformation properties under $T$ of the Majorana fields.} We will also refer to $T$ as a discrete $\mathbb{Z}_2$ R-symmetry, due to its action on the superspace coordinates. In certain cases this is enough to prohibit any term which can lift a moduli space. 

A simple example is $\mathcal{N}=1$ SQED, consisting of a gauge multiplet, a single matter multiplet $\Phi$ and vanishing Chern-Simons level. Classically there is a moduli space parametrized by $|\Phi|$. Gauge invariance forces the superpotential to be a function of $|\Phi|^2$, and so there are no possible superpotential terms which are odd under $\mathbb{Z}_2^R$. As a result a superpotential cannot be generated due to quantum corrections and the moduli space becomes exact quantum-mechanically.\footnote{At least to all orders in perturbation theory. There are cases where non-perturbative corrections can lift the moduli space \cite{Gaiotto:2018yjh}.} Note that adding a Chern-Simons term breaks the R-symmetry and so lifts the moduli space (although in some cases an R-symmetry is emergent in the IR, again leading to a moduli space \cite{Choi:2018ohn,Aharony:2019mbc}). 

In the following we will study another simple example of a Wess-Zumino model that admits a moduli space: the real $ABC$ model. This consists of
three real superfields and superpotential 
\begin{equation}\label{eq_ABC}
    W=\frac{g}{2}ABC\;.
\end{equation}
Classically, this theory has a moduli space consisting of three branches where on each branch only one of the three fields gets a vev. To find which branches are protected from quantum corrections, we analyze the symmetries of the model.
There is a $\mathbb{Z}_2^R$ symmetry that flips the sign of all the superfields $(A,B,C)\to (-A,-B,-C)$, as well as an $S_4$ symmetry group that is obtained composing the obvious $S_3$ permutation group with the action $(A,B,C)\rightarrow(-A,-B,C)$.\footnote{$S_4$ acts as the group of permutations of the set  $\{A+B-C,\,A+C-B,\,C-A+B,\,-A-B-C\}$.}
These are enough to constrain all possible effective action terms to be of the form
\begin{equation}
    W_{\text{eff}}\supset ABCf(A^2,B^2,C^2)\;,
\end{equation}
for some function $f$. Since none of these terms lift the classical moduli space, we find that it is exact quantum-mechanically.

\subsection{\texorpdfstring{$\epsilon$}{epsilon}-expansion for \texorpdfstring{$3d$ $\mathcal{N}=1$}{3d N=1} theories}\label{sec:epsilon_exp}

The coupling $g$ in the $3d$ model~\eqref{eq_ABC} is strongly relevant. In order to obtain a more tractable weakly coupled theory we would then like to analytically continue the real $ABC$ model to $4-\epsilon$-dimensions and use the $\epsilon$-expansion.\footnote{The $ABC$ model and other $\mathcal{N}=1$ SCFTs have also been studied directly in $3d$ via the numerical bootstrap~\cite{Rong:2019qer}.} However, naively there is an obstruction to using this method: $3d$ $\mathcal{N}=1$ consists of two supercharges, while the minimal number of supercharges in $4d$ is four. Equivalently, our theories will include $3d$ Majorana fermions, which have half the number of degrees of freedom as the minimal fermionic representation of the $4d$ Lorentz group. It is thus not immediately clear how to study such theories in the $\epsilon$-expansion from $4d$.

A solution was proposed in \cite{Fei:2016sgs,ThomasSeminar}. Consider a general $3d$ $\mathcal{N}=1$ Wess-Zumino (WZ) theory, consisting of interacting scalar multiplets $\Psi_i$ for $i=1,...,N_\Psi$ whose components are real Bosons $\phi_i$ and Majorana Fermions $\psi_i$. 
In components, the Lagrangian takes the form
\begin{equation}
    \mL=\frac12(\partial_\mu\phi_i)^2-\frac{i}2\psi_i^\alpha\partial_{\alpha\beta}\psi_i^\beta-\frac{i}{2}\partial_i\partial_jW\psi^\alpha_i\psi_{\alpha j}-\frac12(\partial_iW)^2\;,
\end{equation}
where $W(\Psi_i,\bar\Psi_i)$ is the superpotential. 
We instead choose to study a (non-SUSY) theory with an additional flavor index $a=1,...,N_{\psi}$ for the fermions, and Lagrangian
\begin{equation}\label{eq_eps_ep}
     \mL=\frac12(\partial_\mu\phi_i)^2-\frac{i}2\psi_{ai}^{\alpha}\partial_{\alpha\beta}\psi_{ai}^\beta-\frac{i}{2}\partial_i\partial_jW\psi^\alpha_{ai}\psi_{\alpha a j }-\frac12(\partial_iW)^2\;.
\end{equation}
This theory reduces to the SUSY theory when $N_{\psi}=1$. On the other hand, it can be studied in the $\epsilon$-expansion for $N_{\psi}$ a multiple of 2 by repackaging the $3d$ two-component Majorana fermions into $N_{\psi}/2$ $4d$ four-component Majorana fermions. We will go a step further and work with $N_f$ a multiplet of 4, in which case the fermions can be repackaged into $N_{\psi}/4$ $4d$ Dirac fermions. The $4d$ Lagrangian is
\begin{equation}
\label{eq:4d_lagrangian}
\mL=\frac12(\partial_\mu\phi_i)^2+i\bar{\psi}_{bi}\slashed{\pd}\psi_{bi}-\partial_i\partial_jW\bar\psi_{bi}\psi_{ b j }-\frac12(\partial_iW)^2\;,
\end{equation}
where now we sum over $b=1,...,N_{\psi}/4$. The Lagrangian \eqref{eq:4d_lagrangian} can be studied in the $\epsilon$-expansion using standard techniques. Eventually one sets $N_f=1$, which should correspond to the original $3d$ $\mathcal{N}=1$ theory. This strategy was tested for the $3d$ $\mathcal{N}=1$ Ising model in \cite{Fei:2016sgs} and for additional $3d$ $\mathcal{N}=1$ theories in \cite{Liendo:2021wpo,Benini:2018bhk,Benini:2018umh}.

\subsection{The real \texorpdfstring{$ABC$}{ABC} model in \texorpdfstring{$4-\epsilon$}{4-eps} dimensions}

We now consider the continuation of the model~\eqref{eq_ABC} to $4-\epsilon$ dimensions.  Using the prescription explained in~\eqref{eq_eps_ep}, the renormalized Euclidean Lagrangian reads
\begin{equation}\label{eq_ABC_action1}
\begin{split}
\mL &=\frac{1}{2}(\pd a)^2+\frac{1}{2}(\pd b)^2+\frac{1}{2}(\pd c)^2+
\bar{\tilde{a}}_i\slashed{\pd}\tilde{a}_i+\bar{\tilde{b}}_i\slashed{\pd}\tilde{b}_i+
\bar{\tilde{c}}_i\slashed{\pd}\tilde{c}_i
\\ &
+\frac{g \mu^{\epsilon/2}}{2}\left[a\left(\bar{\tilde{b}}_i\tilde{c}_i+\bar{\tilde{c}}_i\tilde{b}_i\right)
+b\left(\bar{\tilde{c}}_i\tilde{a}_i+\bar{\tilde{a}}_i\tilde{c}_i\right)
+c\left(\bar{\tilde{a}}_i\tilde{b}_i+\bar{\tilde{b}}_i\tilde{c}_i\right)\right]
\\&
+\frac{g^2 \mu^{\epsilon}}{8}\left(a^2b^2+b^2 c^2+c^2a^2\right)\,,
\end{split}
\end{equation}
where $\mu$ is the sliding scale, $g$ is the dimensionless renormalized coupling. The fields $a,b,c$ are real scalars while $\tilde{a}_i,\tilde{b}_i,\tilde{c}_i$ are Dirac spinors with $i=1,...,N_\Psi$. We perform calculations with $N_{\psi}$ arbitrary and set $N_{\psi}=1/4$ at the end. By the discussion in section~\ref{sec:epsilon_exp} the model should describe the $3d$ $\mathcal{N}=1$ ABC model~\eqref{eq_ABC}, so that e.g.~$a,\tilde a$ will combine to form the scalar superfield $A$.

As mentioned in section~\ref{subsec_ex_3d_SUSY}, the model is invariant under a discrete internal $S_4$ group which is obtained by composing $S_3$ permutations of the operators with the $\mathds{Z}_2$ flip of the sign of two arbitrary superfields. Additionally, the model is invariant under a discrete chiral symmetry, under which the Dirac fields transform as $(\tilde{a}_i,\tilde{b}_i,\tilde{c}_i)\rightarrow (\gamma_5\tilde{a}_i,\gamma_5\tilde{b}_i,\gamma_5\tilde{c}_i)$, and all the scalars flip sign $(a,b,c)\rightarrow(-a,-b,-c)$.

We work in dimensional regularization within the minimal subtraction scheme, see appendix~\ref{app_ABC_CFT_data} for details. Setting $N_\Psi=1/4$, the beta function of the coupling is 
\begin{equation}\label{eq_ABC_beta_g}
\beta_g=-\frac{\epsilon}{2}g+\frac{5g^3}{4(4\pi)^2}-\frac{9g^5}{8(4\pi)^4}+O\left(\frac{g^7}{(4\pi)^6}\right)\,.
\end{equation}
We see that the model~\eqref{eq_ABC_action1} admits a perturbative fixed point in $4-\epsilon$ dimensions at
\begin{equation}\label{eq_ABC_g_fix}
\frac{g^2_*}{(4\pi)^2}=\frac{2  }{5}\epsilon+\frac{18 }{125}\epsilon ^2
+O\left(\epsilon^3\right)\,.
\end{equation}
We will focus on such fixed-point.  At the fixed-point correlation functions are independent of the sliding scale (provided we rescale the operators appropriately, see~\eqref{eq_app_ABC_wavefunction}). In appendix~\ref{app_ABC_CFT_data} we list some perturbative results for the CFT data that we will need in the next sections. In what follows, we will drop the subscript from the coupling, leaving understood that we always work at the fixed point.  

A technical remark is in order.  As further explained in appendix~\ref{app_ABC_CFT_data}, we work with operators in the minimal subtraction scheme rather than with canonically normalized ones. For instance, the two-point function of $a(x)$ in the conformal vacuum to one-loop order reads
\begin{equation}
    \langle a(x) a(0)\rangle=\frac{\mathcal{N}_{AA}}{|x|^{2\Delta_A}}\,,
\end{equation}
where $\Delta_A=1-\frac{2\epsilon}{5}+O\left(\epsilon^2\right)$ is the scaling dimension (the two-loop result is given in~\eqref{eq_app_DeltaA} of the appendix) and 
\begin{equation}\label{eq_ABC_N_AA}
\mathcal{N}_{AA}=\frac{1}{(d-2)\Omega_{d-1}}\left[1-\frac{\epsilon}{10} \left(2+\gamma +\log \pi \right)+O\left(\epsilon^2\right)\right]\,.
\end{equation}
The canonically normalized operator is given by $\hat{a}\equiv a/\sqrt{\mathcal{N}_{AA}}$.

There are two main technical advantages in working with operators in the minimal subtraction scheme, rather than with canonically normalized ones. First, in this scheme, we do not have to compute explicitly the normalization factors $\mathcal{N}_{ij}$ of the operators exchanged in the OPE; this will reduce the amount of work required in the next section to compare the OPE with explicit results for the two-point functions in the moduli space. Additionally,  OPE coefficients are particularly simple at leading order, since the fusion is trivial at tree-level - e.g. $a(x)\times a(0)=\mathcal{N}_{AA}\mathds{1}/x^2+a^2(0)+\ldots$.

As explained in section~\ref{subsec_ex_3d_SUSY}, we expect that the theory~\eqref{eq_ABC_action1} admits a moduli space of vacua in which one of the fundamental fields acquires a vev. It is instructive to check explicitly that the quantum effective potential admits a flat direction. Without loss of generality we choose to focus on a vacuum such that
\begin{equation}
\langle a\rangle \neq 0\,,\qquad
\langle b\rangle=\langle c\rangle=0\,.
\end{equation}
In this vacuum the internal symmetry is broken to a $\mathds{Z}_2\times\mathds{Z}_2$ subgroup consisting of the exchange of $b$ and $c$ and the transformation $(a,b,c)\rightarrow(a,-b,-c)$. Because of this symmetry we can set both $b$ and $c$ self-consistently to zero, and we find that the one-loop effective potential is given by (note $\mu^\epsilon=1$ to this order)
\begin{align}\nonumber
V_{eff}(a)\vert_{b=c=0}&=2\times\frac12\text{Tr}\left[\log\left(-\pd^2+g^2a^2\right)\right]- N_\Psi\left\{
\text{Tr}\left[\log\left(\slashed{\pd}-ga\right)\right]+
\text{Tr}\left[\log\left(\slashed{\pd}+ga\right)\right]
\right\} \\[0.5em]
&=\left(1-4 N_\Psi\right)\text{Tr}\left[\log\left(-\pd^2+g^2a^2\right)\right]
\stackrel{N_\Psi=1/4}{=}0\,,
\end{align}
which indeed vanishes for the prescribed number of fermions.

The moduli space is parametrized by a single real coordinate, the expectation value $\langle a \rangle$.
As the field $a$ has a non-trivial scaling dimension $\Delta_A = 1 + O(\epsilon)$, it is slightly awkward to work with $\langle a \rangle$. We find it  more convenient to parametrize the moduli space by a coordinate $v$ of dimension one, {\it defined} by the equation 
\begin{equation}\label{eq_vev_a}
    \langle a\rangle\equiv  v^{\frac{d-2}{2}}\left(\frac{gv}{2}\right)^{\Delta_A-\frac{d-2}{2}}=v\left[1+O\left(\epsilon\right)\right]\,.
\end{equation}
It is  redundant but useful to also introduce a mass scale $m$ as
\begin{equation}\label{eq_vev_a_2}
 \frac{g}{2}\langle a\rangle  \equiv 
   m^{\Delta_A} =m\left[1+O\left(\epsilon\right)\right]\,.
\end{equation}
The point of this definition is that $m$ coincides at tree level with the mass of the gapped particles.
Clearly $m$ and $v$ are related as 
\begin{equation}\label{eq_def_m}
m^{\frac{d-2}{2}}\equiv\frac{g v^{\frac{d-2}{2}}}{2}\,.
\end{equation}
These definitions will be convenient in the next section, when we will compare perturbative two-point functions in the moduli space with the OPE.

To compute correlation functions in the moduli space, we simply need to shift $a\rightarrow a+\langle a\rangle$ in the Lagrangian~\eqref{eq_ABC_action1}. Neglecting $O(\epsilon)$ terms and suppressing the fermion flavor index $i$ to avoid clutter of notation, we obtain
\begin{equation}\label{eq_ABC_action2}
\begin{split}
\mL &\simeq\frac{1}{2}(\pd a)^2+\frac{1}{2}\sum_{\pm}\left[(\pd x_\pm)^2+m^2 x_{\pm}^2\right]+\bar{\tilde{a}}\slashed{\pd}\tilde{a}
+\sum_{\pm}\left[\bar{\tilde{x}}_{\pm}\left(\slashed{\pd}\mp m\right)\tilde{x}_{\pm}\right]\\&
-\frac{g}{2}a\left(\bar{\tilde{x}}_{-}\tilde{x}_{-}-\bar{\tilde{x}}_{+}\tilde{x}_{+}\right)-
\frac{g}{2}x_-\left(\bar{\tilde{x}}_{-}\tilde{a}+\bar{\tilde{a}}\tilde{x}_{-}\right)
+\frac{g}{2}x_+\left(\bar{\tilde{x}}_{+}\tilde{a}+\bar{\tilde{a}}\tilde{x}_{+}\right)
\\&
+\frac{g m}{2}a(x_+^2+x_-^2)+\frac{g^2}{8}a^2(x_+^2+x_-^2)+\frac{g^2}{32}(x_+^2-x_-^2)^2\,,
\end{split}
\end{equation}
where we defined the following combinations
\begin{equation}\label{eq_ABC_X_def}
x_{\pm}=\frac{b\pm c}{\sqrt{2}}\,,
\end{equation}
and similarly for the fermions. 
We therefore see that the spectrum consists of a massless superfield, i.e. the dilaton multiplet, and two massive superfields. 

As usual in perturbative theories, the mass $m$ is proportional to the coupling times the expectation value of the field. As we will explain in more detail in the next section, this fact implies that, in order to reproduce correlation functions in the moduli space at arbitrary distances from the OPE, we need to known certain CFT data to arbitrarily high order in $\epsilon$. 
This is because using the action~\eqref{eq_ABC_action2} we formally work in the double-scaling limit $g\rightarrow 0$, $\langle a\rangle\rightarrow\infty$ with $g \langle a\rangle\sim m=\text{fixed}$, in which we resum the effect of infinitely many vertices proportional to $\sim g \langle a\rangle$ with respect to the perturbation theory in the conformal vacuum. The existence of this double-scaling limit can also be understood rescaling all the fields in~\eqref{eq_ABC_action1} by $1/g$, so that the coupling appears only as a $1/g^2$ overall factor in front of the action. Then the definition~\eqref{eq_vev_a_2} ensures that the vev of the rescaled field, and therefore the rescaled action in the moduli space, depend only on $m$ (and $\mu$). Physically, this is just the statement that in perturbation theory we work perturbatively in the coupling, but to all orders in the mass $m\sim g v$ of the particles.

\section{Bootstrap equation in the perturbative \texorpdfstring{$ABC$}{ABC} model}\label{sec_bootABC}

This section contains our main technical results.
We use the real  $ABC$ model in $d  = 4- \epsilon$ dimensions to illustrate how the bootstrap equation works in perturbation theory, up to one-loop order.

\subsection{Tree-level two-point functions}

From the action~\eqref{eq_ABC_action2}, we readily obtain the two-point function of the fundamental field, which is completely specified by
\begin{align}\label{eq_aa_tree}
\langle a(x)a(0)\rangle&=\langle a\rangle^2+\frac{1}{(d-2)\Omega_{d-1}|x|^{d-2}}+O(g^2)\,,\\
\langle b(x) b(0)\rangle&=\langle x_+(x) x_+(0)\rangle=\frac{m^{d-2}}{(2\pi)^{d/2}(m|x|)^{\frac{d-2}{2}}}K_{\frac{d-2}{2}}(m|x|)+O(g^2)\,,
\label{eq_bb_tree}
\end{align}
where we recall that $\Omega_{d-1}=2\pi^{d/2}/\Gamma(d/2)$ is the volume of the $d-1$ dimensional sphere.  To this order, the two-point functions~\eqref{eq_aa_tree} and~\eqref{eq_bb_tree} are saturated by the exchange of, respectively, a massless and a massive particle on the right-hand side of~\eqref{eq_master}.
We will refer to the results~\eqref{eq_aa_tree} and~\eqref{eq_bb_tree} as tree-level, since their connected terms are $O(g^0)$ and are simply given by the free propagator.  On the other hand, the disconnected term is of order $O(g^{-2})$ when working at fixed $m$,
\begin{equation}\label{eq_vev_a_3}
 \langle a\rangle^2=\frac{4}{g^2}m^{\Delta_A}\,,   
\end{equation}
where we used~\eqref{eq_vev_a_2}. As we discussed in section~\ref{subsec_convergence}, it follows from the structure of the free-theory propagator that the OPE converges to this order.

\subsubsection*{Two-point function and OPE at order \texorpdfstring{$O(g^{-2})$}{gminus2}}

In the following we compare~\eqref{eq_aa_tree} and~\eqref{eq_bb_tree} with the OPE expansion~\eqref{eq_OO_OPE} in greater detail. Our motivation for doing so is twofold. First, as we will see, the seemingly trivial results~\eqref{eq_aa_tree} and~\eqref{eq_bb_tree} require several nontrivial identities and cancellations between the anomalous dimensions and OPE coefficients of the operators of the theory when matched to a conformal OPE. Therefore, studying the OPE provides an alternative viewpoint on the necessity of special selection rules for the existence of a moduli space in the double-scaling limit $g\rightarrow 0$ with fixed $m$. Additionally, we would like to understand how the long-distance behaviour of the correlation functions arises from the OPE. We will then generalize these considerations to one-loop order.

Consider first the correlation functions to order $O(1/g^2)$: 
\begin{align}\label{eq_aa_gminus2}
\langle a(x)a(0)\rangle=v^2+O(g^0)\,,\qquad
\langle b(x) b(0)\rangle=\langle c(x) c(0)\rangle=O(g^0)\,.
\end{align}
Physically, to the order $O(1/g^2)$ the correlators are purely classical, and thus are evaluated replacing the fields with their classical expectation value.
Equations~\eqref{eq_aa_gminus2} match on the nose the result of the tree-level OPE:
\begin{align}\nonumber
 a(x)a(0)&=\frac{\mathds{1}}{4\pi x^2}+a^2(0)+\text{descendants}+O\left(g^2\right)\,,\\
 b(x)b(0)&=\frac{\mathds{1}}{4\pi x^2}+b^2(0)+\text{descendants}+O\left(g^2\right)\,, \label{eq_aa_tree_level_OPE}
\end{align}
from which equations~\eqref{eq_aa_gminus2} arise using $\langle a^2\rangle=v^2\left[1+O\left(g^2\right)\right]$ and $\langle b^2\rangle=\langle c^2\rangle=0$.

The simplicity of the matching between the OPE expansion and the explicit results from the moduli space action~\eqref{eq_ABC_action2} is a consequence of the selection rules of the $ABC$ model.
To see this, note that the tree-level vev of the operators $a^n$ is proportional to inverse powers of $g$ in the double scaling limit with fixed $m\sim gv $:
\begin{equation}\label{eq_ABC_an_vev}
    \langle a^n\rangle=v^n=(2m)^n/g^n\,.
\end{equation}
Therefore, a priori, it was possible for the inverse powers of $g$ in~\eqref{eq_ABC_an_vev} to compensate for the factors of the coupling in the higher order terms of the OPE~\eqref{eq_aa_tree_level_OPE} (indeed, we have seen that the $O(g^0)$ OPE yields a $O(g^{-2})$ contribution to the correlation function through the expectation value of $a^2$). This however did not happen in the $ABC$ model due to the specific structure of the interactions.

To make the former discussion more precise, imagine perturbing the action~\eqref{eq_ABC_action1} with interaction vertices of the form $~g^2 a^4$ and $~g^2 a^3 b$. These terms of course destroy the moduli space, but we neglect this fact for now\footnote{These terms might also spoil the conformal invariance of the model at one-loop if included on their own, but for now all we need is that the model is classically conformal.} - we will soon recover this conclusion from the viewpoint of the OPE. It is easy to see that these vertices give rise to OPE coefficients of order $g^{n-2}$ for the operators $a^n$:
\begin{equation}\label{eq_ABC_enhanced_OPEs}
    \delta\mL_{int}\sim \#g^2 a^4+\#g^2 a^3 b\quad\implies\quad
\begin{cases}
a(x)a(0)\supset\sum_{n=\text{even}}\# g^{n-2}x^{n-2}a^n(0)\,,\\
b(x)b(0)\supset \sum_{n=\text{even}} \# g^{n-2}x^{n-2}a^n(0)\,,
\end{cases}
\end{equation}
where we denoted with $\#$ some $O(1)$ factors which are irrelevant for our discussion. Therefore in the presence of such vertices, each term $g^{n-2}a^n$ in the OPE would result in a $\sim g^{n-2}v^n\sim m^n/g^2$ contribution to the correlators~\eqref{eq_aa_gminus2},  since the vev~\eqref{eq_ABC_an_vev} scales as $g^{-n}$. By dimensional analysis, these term would yield a $x$-dependent contribution to the correlation functions already at order $O(g^{-2})$.

Therefore we conclude that it is crucial that the vertices in~\eqref{eq_ABC_enhanced_OPEs} are tuned to zero for the correlators~\eqref{eq_aa_gminus2} to assume a simple factorized form, compatible with $\langle a\rangle=v$, at the classical level. Of course, that these terms must be tuned to zero is not a surprise, since they break supersymmetry and the internal symmetry of the $ABC$ model, that ensure the existence of a flat direction. Our point is simply that the agreement between the conformal OPE and the seemingly trivial expressions~\eqref{eq_aa_gminus2} already at this order relies on the nontrivial structure of the CFT data of the theory.

\subsection{General power counting in perturbative Yukawa models}\label{subsec_power_counting}

In what follows we would like to systematize the comparison between the OPE and the perturbative results in the moduli space at small $g$ and fixed $m$. To this aim, it is useful to make a digression and consider the scaling with the coupling of correlation functions in \emph{generic} weakly coupled theories in a spontaneously broken phase (either conformal or not).

Let us specify what we mean by generic weakly coupled theory. We consider a four-dimensional model consisting of an arbitrary number of scalars and fermions (the inclusion of vector bosons is straightforward) where we allow for all possible mass and renormalizable interaction terms compatible with the Euclidean group. We further assume the most natural scenario, in which the scaling of all terms is set by a single coupling $g$ and a mass term $m$ up to $O(1)$ prefactors. This means that all masses are proportional to $m$, and Yukawa couplings, scalar cubic vertices and scalar quartics scale as, respectively, $ g$, $g m$ and $ g^2$. 

Let us estimate OPE coefficients and one-point functions in this scenario.
Naturally, the vev $v$ of fundamental scalar fields is set by $m/g$. 
Let us generically denote with $\phi$ the bosonic fundamental operators and with $\psi$ the fermionic ones.  A generic composite operator is made out of products of the fundamental fields and derivatives, schematically
\begin{equation}
\mO_{n_{\phi},n_{\psi},n_{\pd}}=\pd^{n_{\pd}}\phi^{ n_{\phi}}\psi^{n_{\psi}}\,.
\end{equation}
It is easy to get convinced that the vev of a {\it generic} such operator in the limit of scall coupling and fixed mass is parametrically bounded as
\begin{equation}\label{eq_counting_vev}
\langle \mO_{n_{\phi},n_{\psi},n_{\pd}}\rangle\lesssim  v^{n_{\phi}}m^{\frac32 n_{\psi}+n_{\pd}}\sim  m^{n_{\phi}+\frac32 n_{\psi}+n_{\pd}} /g^{n_{\phi}}\,.
\end{equation}
For operators with no derivatives or fermions $n_{\pd}=n_{\psi}=0$,~\eqref{eq_counting_vev} agrees with~\eqref{eq_ABC_an_vev}. We also used that fermionic bilinears acquire a vev at loop level via a self-contraction, e.g. $\langle\bar{\psi}\psi\rangle\sim m^3$. Finally derivatives kill the classical expectation value, but when acting on propagators in loop diagrams also yield factors $m$.

Similarly, connected three-point functions of two fundamental scalar fields and a composite operator scale at most as an amputated $2\rightarrow n_{\phi}+n_{\psi}$ correlator, therefore we have the following parametric bound
\begin{equation}\label{eq_counting_OPE}
\langle \phi\phi \mO_{n_{\phi},n_{\psi},n_{\pd}}\rangle_c\lesssim g^{n_{\phi}+n_{\psi}-2}\,,
\end{equation}
up to a function of $m$ and the distance between the operators (which is trivial for $m=0$).

Of course,~\eqref{eq_counting_vev} and~\eqref{eq_counting_OPE} are just crude upper bounds; for instance, spinning operators or total derivatives always have trivial expectation value. For our purposes it is enough to note that at fixed and sufficiently large classical scaling dimension, in a generic model we find a nonzero number of operators for which the inequalities~\eqref{eq_counting_vev} and~\eqref{eq_counting_OPE} are saturated. 
There are however two nontrivial exceptions to these estimates. One concerns  operators without fermions but at least two derivatives. In this case, since for the operator to acquire a vev we need to self-contract at least two scalars via a propagator, we find
\begin{equation}\label{eq_counting_vev_ex}
\langle \mO_{n_{\phi},0,n_{\pd}}\rangle\lesssim v^{n_{\phi}-2}m^{n_{\pd}+2}\sim m^{n_{\phi}+n_{\pd}} /g^{n_{\phi}-2}\qquad\text{for }\;n_{\pd}>0\,,
\end{equation}
which is $g^2$ smaller than~\eqref{eq_counting_vev}. The other exception concerns the three-point functions of two fundamental scalar fields and the identity operator, formally an operator of the form $\mO_{0,0,0}$, in which case $\langle \phi\phi \mathds{1}\rangle_c\sim O(1)$. For this reason, in what follows we will treat the contribution of the identity to the OPE separately from the rest.

Given~\eqref{eq_counting_vev} and~\eqref{eq_counting_OPE}, we can now readily write the generic expansion of CFT data and one-point functions in generic Yukawa models in the $\epsilon$-expansion, like the $ABC$ model considered here.  The scaling dimensions admit the usual expansion:
\begin{equation}\label{eq_Delta_general_expansion}
\Delta_{\mO_{n_{\phi},n_{\psi},n_{\pd}}}=n_{\phi}+\frac32 n_{\psi}+n_{\pd}+\epsilon\,\gamma_{\mO}^{(1)}+\epsilon^2\,\gamma_{\mO}^{(2)}+\ldots\,.
\end{equation}
Note that here we define anomalous dimensions including the \emph{classical} piece $\frac{d-2}{2}n_{\phi}+\frac{d-1}{2}n_{\psi}-n_{\phi}-\frac32 n_{\psi}$.  Identifying $g^2\sim \epsilon$,  OPE coefficients are expanded as
\begin{equation}\label{eq_OPE_general_expansion}
g_{\phi\phi\mO_{n_{\phi},n_{\psi},n_{\pd}}}=\epsilon^{\frac{n_{\phi}+n_{\psi}}{2}-1}\left[g_{\phi\phi\mO}^{(0)}+\epsilon \,g_{\phi\phi\mO}^{(1)}+\epsilon^2 \,g_{\phi\phi\mO}^{(2)}+\ldots\right]\,.
\end{equation}
Finally, the vev of a composite operator reads
\begin{equation}\label{eq_1pt_general_expansion}
\langle \mO_{n_{\phi},n_{\psi},n_{\pd}}\rangle =
\frac{m^{n_{\phi}+\frac32 n_{\psi}+n_{\pd}}}{\epsilon^{n_{\phi}/2}}\left[\xi_{\mO}^{(0)}+\epsilon\,\xi_{\mO}^{(1)}+\epsilon^2\,\xi_{\mO}^{(2)}+\ldots\right]\,.
\end{equation}
Note that, since $\langle \mO\rangle\propto m^{\Delta_{\mO}}$ by dimensional analysis,  the coefficients $\xi_{\mO}^{(k)}$ of the expansion in~\eqref{eq_1pt_general_expansion} must contain powers of $\log m$.

It is important that the structure of these expansions is not affected by operator mixing. For instance, an operator $\sim \phi^{2n}$ may mix with $\sim g^2(\pd\phi)^2\phi^{n-4}$, which contributes at subleading order to the vev~\eqref{eq_1pt_general_expansion} and at leading order to the OPE~\eqref{eq_OPE_general_expansion}.

\subsection{Consistency conditions at tree level}
We now proceed with a detailed analysis of how the consistency conditions are obeyed in the real $ABC$ model at tree level.

\subsubsection*{Order \texorpdfstring{$O(g^{-2})$}{gminus2}}

Of course, the real $ABC$ model is not a completely generic model, and its observables obey other selection rules besides Poincar\'e invariance. In particular, only superconformal primaries can acquire a vev. Additionally, the internal symmetry group constrains, besides others, the structure of the OPE and one-point functions, and in particular implies that the leading term in~\eqref{eq_OPE_general_expansion} vanishes in many cases. We have already seen that some of these cancellations are also required for the OPE to match the trivial results~\eqref{eq_aa_tree} and~\eqref{eq_bb_tree} at order $O(g^{-2})$ - below we extend the discussion to order $O(g^0)$. We will discuss the order $O(g^2)$ in section~\ref{sec_ABC_1_loop}.

Let us first reproduce our former discussion of~\eqref{eq_aa_gminus2} in the current notation. Using~\eqref{eq_counting_vev_ex}, we see that at order $O(\epsilon^{-1})$ only operators of the form $\mO_{2n,0,0}=\phi^{2n}$ (with $\phi=a,b,c$) may potentially contribute to the two-point function of the fundamental field (odd powers are excluded because of the internal $\mathds{Z}_2$). Therefore we must have
\begin{align}\label{eq_ABC_aa_crossing_OPE_gminus2}
v^2&=\sum_{n \in \mathbb{N}}\sum_{\mO_{2n,0,0}}
\frac{ m^{2n}/\epsilon}{|x|^{2-2n}}\xi_{\mO}^{(0)}g_{AA\mO}^{(0)}\,,\\
0&=\sum_{n \in \mathbb{N}}\sum_{\mO_{2n,0,0}}
\frac{ m^{2n}/\epsilon}{|x|^{2-2n}}\xi_{\mO}^{(0)}g_{BB\mO}^{(0)}\,.
\end{align}
Note that the identity does not contribute at this order.
As we have seen before, these equations are satisfied because the only operators acquiring a vev are of the form $a^{2n}$. These operators have an anomalously small OPE coefficient because supersymmetry and the internal group forbid $g^2 a^4$ and $g^2 a^3 b$ interactions in the action, hence implying $g_{AA A^{2n}}^{(0)}=g_{BB A^{2n}}^{(0)}=0$.

Beyond leading order additional operators acquire a vev, and we expect that all scalar superconformal primary operators, up to the requirements of the internal symmetry, acquire an expectation value at sufficiently high order in perturbation theory. Additionally, the operators that diagonalize the dilation operator in the interacting theory are nontrivial linear combinations of $A^{2n}$ and other classically degenerate operators, see appendix~\ref{app_ABC_CFT_data}. For these reasons,  the comparison with the OPE is less trivial at order $O(\epsilon^0)$. We discuss some detailed checks below.

\subsubsection*{Order \texorpdfstring{$O(g^0)$}{g0}}

Let us consider for concreteness the $\langle a a\rangle$ two-point function.  At this order the OPE also receives contributions from operators with an arbitrary number of derivatives and up to two fermionic fields: $\mO_{2n-2k,0,2k}=\pd^{2k}\phi^{2n-2k}$ and $\mO_{2n-3-2k,2,2k}=\pd^{2k}\phi^{2n-3-2k}\bar{\psi}\psi$. Operators of the form $\phi^{2n}\psi^{2}$ in principle could also contribute at this order, but their OPE coefficients vanish because of the discrete chiral symmetry mentioned below~\eqref{eq_ABC_action1}. Therefore we find
\begin{equation}\label{eq_ABC_AA_crossing_OPE_g0}
\begin{split}
\frac{1}{4\pi^2|x|^{2}}+2m^2\xi_A^{(0)}\xi_A^{(1)}
&=\frac{\mathcal{N}_{AA}\vert_{\epsilon^0}}{x^{2}}+\sum_{n\in\mathbb{N}^+}
\frac{m^{2n}\log x^2}{2|x|^{2-2n}}\sum_{\mO_{2n,0,0}}\xi_{\mO}^{(0)}g_{AA\mO}^{(0)}\left(\gamma_{\mO}^{(1)}-2\gamma_{A}^{(1)}\right)\\
&+\sum_{n \in \mathbb{N}}\frac{m^{2n}}{|x|^{2-2n}}
\left\{ \sum_{\mO_{2n,0,0}}\left[\xi_{\mO}^{(1)}g_{AA\mO}^{(0)}+\xi_{\mO}^{(0)}g_{AA\mO}^{(1)}\right]
\right.\\
&\left.
+\sum_{k=0}^{n-2}\sum_{\mO_{2n-3-2k,2,2k}}
\xi_{\mO}^{(0)}g_{AA\mO}^{(0)}+\sum_{k=1}^{n-1}\sum_{\mO_{2n-2k,0,2k}}
\xi_{\mO}^{(1)}g_{AA\mO}^{(0)}\right\}\,,
\end{split}
\end{equation}
where $\mathcal{N}_{AA}$ is given in~\eqref{eq_ABC_N_AA}.
A similar equation holds for the $\langle b b\rangle$ two-point function, with the left hand-side replaced by the short distance expansion of the $4d$ massive scalar propagator~\eqref{eq_scalar_prop}:
\begin{equation}
G_{m^2}(x^2)=\frac{1}{4 \pi ^2 x^2}+\frac{m^2 }{16 \pi ^2}\left[\log \left(\frac{m^2 x^2}{4}\right)+2 \gamma -1\right]+
\frac{m^4 x^2 }{256 \pi ^2}\left[2\log\left(\frac{m^2 x^2}{4}\right)+4 \gamma -5\right]+\ldots\,.
\end{equation}

Let us unpack~\eqref{eq_ABC_AA_crossing_OPE_g0}.  The first term on the left-hand side is the massless scalar propagator, while the second term arises from the expansion of $\langle A\rangle^2$. From~\eqref{eq_vev_a_2} we find:
\begin{equation}
\xi_A^{(0)}=\frac{\sqrt{\frac{5}{2}}}{2 \pi }\,,\qquad
\xi_A^{(1)}=-\frac{ 20 \log m+9}{20 \sqrt{10} \pi }\,.
\end{equation}
As formerly commented,  we work with non-canonically normalized operators, and the dependence on $m$ is fixed by the scaling dimension.
The first term on the right-hand side of~\eqref{eq_ABC_AA_crossing_OPE_g0} is the contribution from the identity in the OPE. The second term arises from the anomalous dimensions of $A$ and the composite operators on the right hand side of~\eqref{eq_ABC_aa_crossing_OPE_gminus2}. On the second and third lines we grouped the first terms in the expansion of the vev and the OPE of the operators $\mO_{2n,0,0}\sim \phi^{2n}$, which already appeared in the previous order, with the first contribution of operators with either derivatives or fermions; note that because of the property~\eqref{eq_counting_vev_ex} the operators $\mO_{2n-2k,0,2k}=\pd^{2k}\phi^{2n-2k}$ have $\xi_{\mO}^{(0)}=0$.

Let us now see explicitly how the leading terms at short distances are obtained from the OPE. The $\sim 1/x^2$ on the left-hand side of the tree-level results is trivially reproduced by the contribution of the identity. The $\sim x^0$ contributions instead arise from operators of the form $ \phi^2$. As detailed in appendix~\ref{app_ABC_CFT_data}, these can be grouped into a singlet $S=A^2+B^2+C^2$ and a doublet $\{D_1,D_2\}=\{A^2-B^2,(A^2+B^2-2C^2)/\sqrt{3}\}$ under $S_4$.  The internal symmetry implies
\begin{equation}
\langle D_1\rangle=\sqrt{3}\langle D_2\rangle\,,\qquad
g_{AA D_1}=-g_{BB D_1}=\sqrt{3}g_{AAD_2}=\sqrt{3}g_{BBD_2}\equiv g_{AAD}\,,
\end{equation}
as well as $g_{AA S}=g_{BBS}$. Therefore, comparing the OPE with the tree-level results for the two-point functions we find
\begin{align}\label{eq_OPEcheck_tree_log1}
0&=\xi_{S}^{(0)}g_{AA S}^{(0)}\left(\gamma_{S}^{(1)}-2\gamma_{A}^{(1)}\right)+\frac{4}{3}\xi_{D}^{(0)}g_{AA D}^{(0)}\left(\gamma_{D}^{(1)}-2\gamma_{A}^{(1)}\right)\, ,\\
\frac{1}{8\pi^2}&=\xi_{S}^{(0)}g_{AA S}^{(0)}\left(\gamma_{S}^{(1)}-2\gamma_{A}^{(1)}\right)
-\frac{2}{3}\xi_{D}^{(0)}g_{AA D}^{(0)}\left(\gamma_{D}^{(1)}-2\gamma_{A}^{(1)}\right)
\,,
\label{eq_OPEcheck_tree_log2}
\end{align}
from the $\log x^2$ terms, as well as
\begin{align}\label{eq_OPEcheck_tree_x0_1}
2\xi_A^{(0)}\xi_A^{(1)}&=\xi_{S}^{(1)}g_{AA S}^{(0)}+\xi_{S}^{(0)}g_{AA S}^{(1)}+\frac{4}{3}\left[
\xi_{D}^{(1)}g_{AA D}^{(0)}+\xi_{D}^{(0)}g_{AA D}^{(1)}\right]\,,\\
\frac{1 }{16 \pi ^2}\left[\log \left(\frac{m^2}{4}\right)+2 \gamma -1\right]&=\xi_{S}^{(1)}g_{AA S}^{(0)}+\xi_{S}^{(0)}g_{AA S}^{(1)}-\frac{2}{3}\left[
\xi_{D}^{(1)}g_{AA D}^{(0)}+\xi_{D}^{(0)}g_{AA D}^{(1)}\right]\,,
\label{eq_OPEcheck_tree_x0_2}
\end{align}
from the $x^0$ pieces. 

To illustrate the potential uses of these relations, note that the tree-level OPEs and vevs are
\begin{equation}
g_{AAS}^{(0)}=\frac{1}{3}\,,\qquad g_{AAD}^{(0)}=\frac{1}{2}\,,\qquad
\xi_S^{(0)}=\xi_D^{(0)}=\frac{5}{8 \pi ^2}\,.
\end{equation}
Using these,~\eqref{eq_OPEcheck_tree_log1}, \eqref{eq_OPEcheck_tree_log2}, \eqref{eq_OPEcheck_tree_x0_1} and~\eqref{eq_OPEcheck_tree_x0_2}, yield relations between the one-loop CFT data of the theory from a tree-level calculation on the moduli space. For instance, if we are given the one-loop anomalous dimensions of $A$,
\begin{equation}
\gamma_A^{(1)}=-\frac{2}{5}\,,
\end{equation}
from~\eqref{eq_OPEcheck_tree_log1} and \eqref{eq_OPEcheck_tree_log2} we can solve for the anomalous dimensions of $S$ and $D$
\begin{equation}
\gamma_S^{(1)}=-\frac{2}{5}\,,\qquad
\gamma_D^{(1)}=-1\,,
\end{equation}
which are in perfect agreement with the results~\eqref{eq_app_DeltaS} and~\eqref{eq_app_DeltaD} from a direct calculation in the appendix. Similarly, from~\eqref{eq_OPEcheck_tree_x0_1} and~\eqref{eq_OPEcheck_tree_x0_2} we obtain the following relation between OPEs and one-point functions
\begin{align}\label{eq_ABC_tree_OPE_1pt_1}
 g_{AAS}^{(1)}+\frac{8\pi^2}{15}\xi_S^{(1)}&= 
 \frac{1}{75} \left[10 \gamma -14-\log (1024)-10 \log m\right]\,,\\
 \label{eq_ABC_tree_OPE_1pt_2}
  g_{AAD}^{(1)}+\frac{4\pi^2}{5}\xi_D^{(1)}&= 
 \frac{1}{100} \left[\log (1024)-10 \gamma -13-50 \log m\right]\,.
\end{align}
These are satisfied by the results~\eqref{eq_app_OPE} and~\eqref{eq_app_1pt_pre} in the appendix, that give
\begin{align}
g_{AAS}^{(1)}&=\frac{1}{15} \left(\log \pi+\gamma  \right)\,,\\
g_{AAD}^{(1)}&=-\frac{1}{20} \left(\log \pi+\gamma  \right)\,,\\
    \xi_S^{(1)}&=-\frac{10 \log m-5 \gamma +14+5 \log (4 \pi )}{40 \pi ^2}\,,\\
    \xi_D^{(1)}&=\frac{5\log(4\pi)-50 \log m-5 \gamma -13}{80 \pi ^2}\,.
\end{align}

It is increasingly difficult to check explicitly the identities arising from higher powers of $x^2$ in~\eqref{eq_ABC_AA_crossing_OPE_g0}, since we need to compute the CFT data of higher dimensional operators including OPE coefficients to very high order in $\epsilon$. As an illustration, from the $x^2\log x^2$ terms of the OPE expansion in~\eqref{eq_ABC_AA_crossing_OPE_g0} we find the following
\begin{align}\nonumber
0&=\xi_{\Sigma}^{(0)}g_{AA \Sigma}^{(0)}\left(\gamma_{\Sigma}^{(1)}-2\gamma_{A}^{(1)}\right)+
\xi_{K}^{(0)}g_{AA K}^{(0)}\left(\gamma_{K}^{(1)}-2\gamma_{A}^{(1)}\right)\\
\label{eq_OPEcheck_tree_x2log1}
&+\frac{4}{3}\xi_{M}^{(0)}g_{AA M}^{(0)}\left(\gamma_{M}^{(1)}-2\gamma_{A}^{(1)}\right)
+\frac{4}{3}\xi_{N}^{(0)}g_{AA N}^{(0)}\left(\gamma_{N}^{(1)}-2\gamma_{A}^{(1)}\right)
\, ,\\  \nonumber
\frac{1}{64 \pi ^2}&=\xi_{\Sigma}^{(0)}g_{AA \Sigma}^{(0)}\left(\gamma_{\Sigma}^{(1)}-2\gamma_{A}^{(1)}\right)+
\xi_{K}^{(0)}g_{AA K}^{(0)}\left(\gamma_{K}^{(1)}-2\gamma_{A}^{(1)}\right)\\
&-\frac{2}{3}\xi_{M}^{(0)}g_{AA M}^{(0)}\left(\gamma_{M}^{(1)}-2\gamma_{A}^{(1)}\right)
-\frac{2}{3}\xi_{N}^{(0)}g_{AA N}^{(0)}\left(\gamma_{N}^{(1)}-2\gamma_{A}^{(1)}\right)
\,,
\label{eq_OPEcheck_tree_x2log2}
\end{align}
where at this order we receive contributions from two singlets, $\Sigma$ and $K$, and two doublets, $M_i$ and $N_i$, see app.~\ref{app_ABC_CFT_data} for the expressions in terms of fundamental fields. Eq.s~\eqref{eq_OPEcheck_tree_x2log1} and~\eqref{eq_OPEcheck_tree_x2log2} are in agreement with the results~\eqref{eq_app_Delta4_singlet},~\eqref{eq_app_Delta4_doublet},~\eqref{eq_app_OPE} and \eqref{eq_app_vev_dim4} in the appendix.

In summary,  already at tree-level the correlators~\eqref{eq_aa_tree} and~\eqref{eq_bb_tree} contain a wealth of information on the CFT data of the theory. The comparison with the OPE shows that the existence of a moduli space relies on nontrivial relations between the CFT data of the theory. To further illustrate this point, in the next section we will extract the $O(\epsilon^2)$ corrections to the OPE coefficients of $S$ and $D$ from the one-loop result for the two-point functions of $A$ and $B$ in the moduli space.

We speculate that the power-counting analysis of this section should be easily generalizable to non conformal theories, with the main difference that the OPE is not constrained by conformal invariance. It might be possible to perform a detailed comparison between perturbative correlators and the OPE in massive theories along the lines of our discussion, perhaps using conformal perturbation theory to evaluate corrections to the OPE. This analysis might prove useful in understanding the properties and structure of the OPE in generic massive theories.\footnote{See~\cite{Hollands:2011gf,Hollands:2023txn} and references therein for discussions of the OPE in massive QFTs.}

\subsection{Two-point functions and consistency relations at one-loop}\label{sec_ABC_1_loop}

In this subsection we extend our considerations to one-loop order.  We will consider both the convergence properties of the short and long distance expansions, as well as some of the consistency relations that arise from the comparison of the explicit results with the OPE.

\begin{figure}[t]
   \centering
		\subcaptionbox{  \label{fig:AA_bubble}}
		{\includegraphics[width=0.23\textwidth]{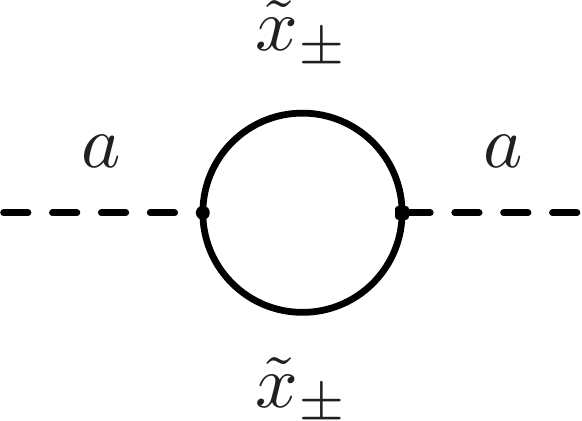}}
 \hspace*{1.5cm}
		\subcaptionbox{ \label{fig:AA_bubbleS}}
		{\includegraphics[width=0.23\textwidth]{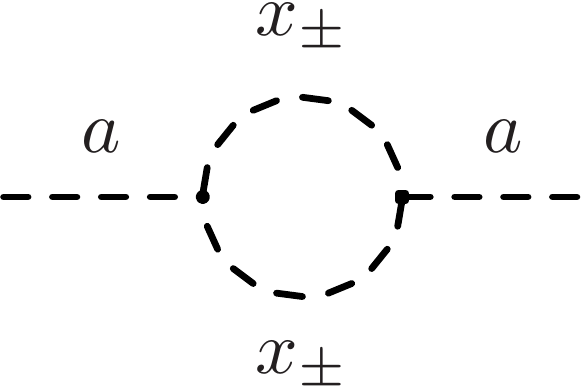}}
		 \hspace*{1.5cm}
  \subcaptionbox{ \label{fig:AA_tadpole}}
		{\includegraphics[width=0.23\textwidth]{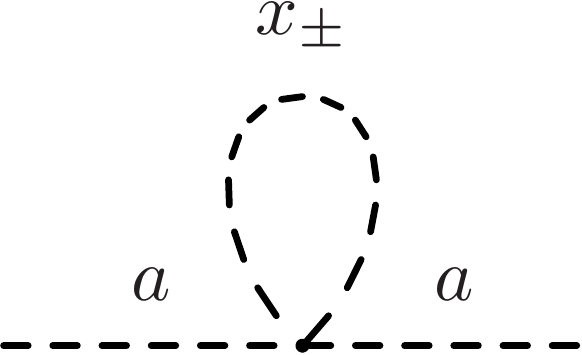}}
		  \\[1em]
		\subcaptionbox{  \label{fig:XX_bubble}}
		{\includegraphics[width=0.23\textwidth]{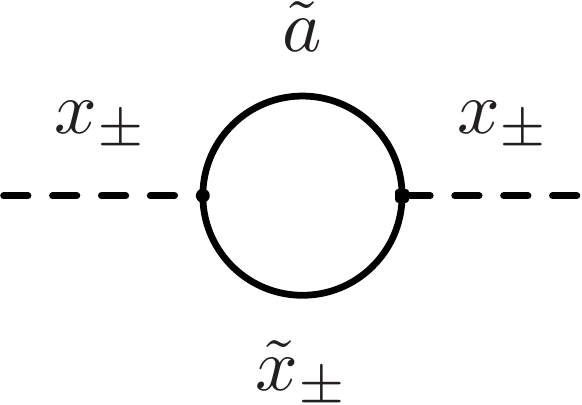}}
 \hspace*{1.5cm}
		\subcaptionbox{ \label{fig:XX_bubbleS}}
		{\includegraphics[width=0.23\textwidth]{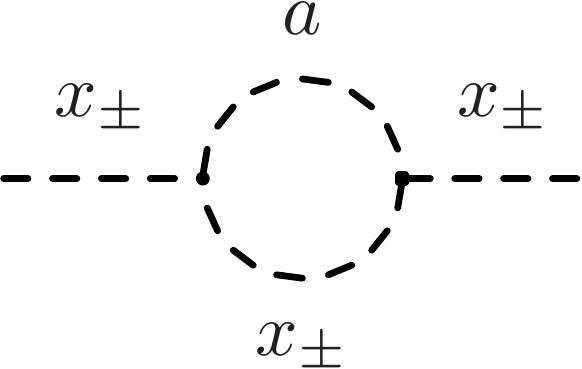}}
		 \hspace*{1.5cm}
  \subcaptionbox{ \label{fig:XX_tadpole}}
		{\includegraphics[width=0.23\textwidth]{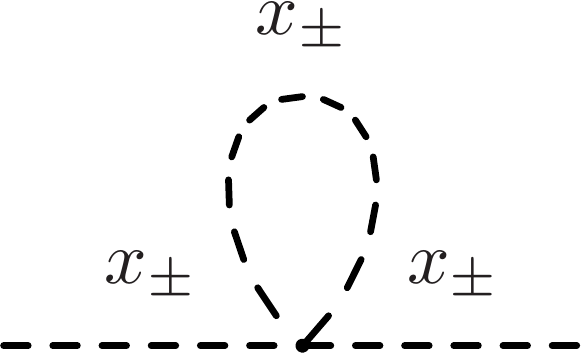}}
        \caption{Diagrams contributing the two-point functions of $a$ and $x_{\pm}$ at one-loop. We neglected diagrams proportional to massless tadpole that vanish identically in dimensional regularization.}
\label{fig:DiagramsLoop}
\end{figure}

The one-loop two-point functions are more easily expressed in momentum space. We define the 1PI factors as
\begin{align}\label{eq_aa2pt_1loop_mom}
\int d^dx e^{ipx}\langle a(x)a(0)\rangle^{(1-loop)}&=\frac{g^2}{(4\pi)^2}\frac{\Gamma_{aa}(p)}{p^4}\,,\\
\label{eq_bb2pt_1loop_mom}
\int d^dx e^{ipx}\langle x_+(x)x_+(0)\rangle^{(1-loop)}&=\frac{g^2}{(4\pi)^2}\frac{\Gamma_{xx}(p)}{(p^2+m^2)^2}\, .
\end{align}
The relevant diagrams are shown in figure~\ref{fig:DiagramsLoop}.  As these show, at one-loop order we introduce the contribution of two-particle states in the spectral density on the right hand side of the bootstrap equation~\eqref{eq_master}. A textbook calculation gives
\begin{align}\label{eq_Gamma_aa}
\Gamma_{aa}(p)&=\frac12 p\sqrt{4 m^2+p^2} \tanh ^{-1}\left(\frac{p}{\sqrt{4 m^2+p^2}}\right)+\frac{p^2}{4}
\left[\log \left(\frac{m^2}{4 \pi }\right)+\gamma -2\right]
\,,\\ \nonumber
\Gamma_{xx}(p)&=\frac{1}{4} p^2 \left[\log \left(\frac{p^2+m^2}{4\pi}\right)+\gamma -2\right]-\frac{3 m^4 }{4 p^2}\log \left(1+\frac{p^2}{m^2}\right)\\ \label{eq_Gamma_bb}
&+\frac{1}{4} m^2 \left[6+3 \log (4 \pi )-3\gamma-2 \log \left(p^2+m^2\right)-2 \log \left(m^2\right)\right]\,;
\end{align}
note that $\Gamma_{aa}(0)=0$, ensuring that the dilaton remains massless also at one-loop level. 

\subsubsection*{Convergence of the short and large distance expansions}

The convergence properties of the low and large momentum expansion follow from the analytic structure of the results.
The 1PI factors~\eqref{eq_Gamma_aa} and~\eqref{eq_Gamma_bb} have a branch cut starting, respectively, at  at $p^2=-4m^2$, due to the continuum of two massive particles (from the diagrams~\ref{fig:AA_bubble} and~\ref{fig:AA_bubbleS}), and at $p^2=-m^2$ (from the diagrams~\ref{fig:XX_bubble} and~\ref{fig:XX_bubbleS}), due to the exchange of a dilaton and a massive particle.  At this order both correlators are analytic for $|p^2|< m^2$, because there is no cut corresponding to the exchange of two or more massless particles. Therefore the small momentum expansion for $|p^2|/m^2<1$ is convergent.  Less obviously, the momentum space OPE expansion is convergent for momenta larger than the branch points, i.e. for $|p^2|/(4m^2)>1$ in~\eqref{eq_Gamma_aa} and for $|p^2|/m^2>1$ in~\eqref{eq_Gamma_bb}.

Now we would like to compare explicitly our results with the OPE as in the previous section. To this aim, we need the correlation functions in position space.  The Fourier transform of~\eqref{eq_bb2pt_1loop_mom} can be expressed in closed form in terms of Bessel functions as
\begin{equation}\label{eq_xx_2pt_1loop}
\begin{split}
\langle x_+(x)x_+(0)\rangle^{(1-loop)}=&-\frac{g^2}{(4\pi)^2}\left\{
    \frac{m K_1(m x)}{16 \pi ^2 x} \left[\log \left(\frac{\pi}{4}   m^2  x^4\right)+3 \gamma +2\right]\right.
    \\
    &\left.+\frac{K_0(m x) }{16 \pi ^2 x^2}\left[2-m^2 x^2 \left(4+2 \log (4 \pi )-5 \log m-2 \gamma \right)\right]
    \right\}\,.
    \end{split}
\end{equation}
We were only able to express the Fourier transform of~\eqref{eq_aa2pt_1loop_mom} as a series expansion for small $x^2$:
\begin{equation}\label{eq_aa_2pt_1loop}
\langle a(x) a(0)\rangle^{(1-loop)}=\frac{g^2}{(4\pi)^2}\sum_{k=-1}^{\infty}x^{2k}m^{2k+2}\left[\alpha_k\log^2|x|+\beta_k\log|x|+\gamma_k\right]\,.
\end{equation}
We obtained closed expressions for all the $\alpha_k$'s and $\beta_k$'s, and we computed the values of the first few $\gamma_k$'s; we will derive below also the asymptotic form of the $\gamma_k$'s for large $k$. A similar expansion can be obtained for the two-point function of $b$ from~\eqref{eq_xx_2pt_1loop}. We provide details of the derivations of~\eqref{eq_xx_2pt_1loop} and~\eqref{eq_aa_2pt_1loop} and the explicit values of the coefficients in app.~\ref{app_Fourier}.

We find again that the OPE is uniformly convergent for any $x^2>0$. For the two-point function of $x_+$, this follows from~\eqref{eq_xx_2pt_1loop} and the properties of the Bessel functions. To see that the OPE converges also for the $\langle a(x) a(0)\rangle$ correlator, we use that the coefficients of the logarithmic terms in the expansion~\eqref{eq_aa_2pt_1loop} of $\langle a(x) a(0)\rangle$ decay faster than exponentially
\begin{equation}\label{eq_ABC_alpha_beta}
    \alpha_k\sim\frac{1}{16 \pi^{5/2} k^{5/2} (k!)^2}\,,\quad
    \beta_k\sim-\frac{\log (k/m)}{8 \pi ^{5/2} k^{5/2} (k!)^2}\quad\text{for}\quad k\rightarrow\infty\,,
\end{equation}
as can be verified from the explicit result in~\eqref{eq_app_alpha_beta} of the appendix. It follows that the $\gamma_k$'s obey a similar behavior. This is because $\langle a(x)a(0)\rangle$ does not grow exponentially at large $x^2$, while~\eqref{eq_ABC_alpha_beta} imply that the sum of the first two terms in~\eqref{eq_aa_2pt_1loop} yield an exponentially growing behaviour $\sim e^{m|x|}$.  Therefore the sum over the $\gamma_k$'s must compensate for that. Evaluating the sum~\eqref{eq_aa_2pt_1loop} via the saddle-point approximation we find that this is possible only if
\begin{equation}
\gamma_k\sim\frac{\log^2(k/m)}{16 \pi^{5/2} k^{5/2} (k!)^2}\quad\text{for}\quad k\rightarrow\infty\,.
\end{equation}
Similarly to the discussion below~\eqref{eq_ABC_Bessel_series}, these observations imply that the long distance limit of the sum in~\eqref{eq_aa2pt_1loop_mom}, which is proportional to $\sim 1/|x|^2$, arises as a consequence of several nontrivial cancellations between the individual contributions in the OPE.

The cuts in momentum space imply that at this order the long distance limit of the dilaton two-point function receives exponentially suppressed contributions $\sim e^{-2m|x|}$. Therefore we conclude that the long distance expansion in coordinate space is asymptotic for both correlators, as expected.

\subsubsection*{Consistency relations at one-loop}

Finally, we can add the one-loop results to the $O(\epsilon)$ terms from the expansion of the tree-level and compare explicitly to the OPE. The check is largely analogous to the one detailed at tree-level. The main difference is that at this order we also have terms proportional to $\log^2|x|$ times the square of the anomalous dimensions of the operators of the form $\phi^{2n}$, 
\begin{equation}
\langle a(x)a(0)\rangle\supset  \epsilon \,m^2\sum_{\mO_{2,0,0}}\xi_{\mO}^{(0)}\lambda_{AA\mO}^{(0)}\left(\gamma_{\mO}^{(1)}-2\gamma_{A}^{(1)}\right)^2\frac{\log^2|x|}{2}\,,
\end{equation}
and similarly for $\langle b(x) b(0)\rangle$. Additionally at this order there are contributions from operators with four fermionic fundamental fields, that did not appear before.

We verified explicitly the agreement with the OPE for the operators of engineering dimension $2$, i.e. $S$ and $D_i$. In particular, we found that the $\epsilon\,x^0\log^2|x|$ and $\epsilon\,x^0\log|x|$ terms are correctly reproduced by the OPE using the results in app.~\ref{app_ABC_CFT_data} for the one- and two-loop anomalous dimensions, and for the one-loop OPE coefficients and one-point functions of $S$ and $D_i$. Additionally, analogously to~\eqref{eq_ABC_tree_OPE_1pt_1} and~\eqref{eq_ABC_tree_OPE_1pt_2}, from the $\epsilon\,x^0$ term we find a relation between the two-loop one-point functions and OPE coefficients of $S$ and $D_i$. From this relation and the results for the one-point functions in~\eqref{eq_app_1pt_S} and~\eqref{eq_app_1pt_D}, we extract the $O(\epsilon^2)$ corrections to the OPE coefficients of $S$ and $D$:
\begin{align}\nonumber
g_{AAS}^{(2)}&=\frac{\log ^2\pi}{150}+\frac{\log \pi }{750} (10 \gamma -17)+
\frac{60 \gamma ^2-204 \gamma+25 \pi ^2 -390}{9000} \,, \\
g_{AAD}^{(2)}&=\frac{\log ^2\pi }{400}+\frac{\log \pi }{1000}(5 \gamma +2)+
\frac{30 \gamma^2+  24\gamma-25 \pi ^2+480}{12000}\,.
\end{align}
Note that a direct calculation of these OPE coefficients would involve rather intricate three-point two-loop integrals. Here instead we extracted the result from a textbook one-loop calculation of a two-point function and from the two-loop one-point functions (whose calculation is simple because of the trivial kinematics).

Finally,  matching the $\epsilon\,x^2\log^2|x|$ terms in the expansion of the two-point functions with the OPE, we find the following conditions:
\begin{align}\nonumber
\frac{1}{320 \pi ^2}&=\xi_{\Sigma}^{(0)}g_{AA \Sigma}^{(0)}\frac{\left(\gamma_{\Sigma}^{(1)}-2\gamma_{A}^{(1)}\right)^2}{2}+
\xi_{K}^{(0)}g_{AA K}^{(0)}\frac{\left(\gamma_{K}^{(1)}-2\gamma_{A}^{(1)}\right)^2}{2}\\
\label{eq_OPEcheck_1loop_x2log1}
&+\frac{4}{3}\xi_{M}^{(0)}g_{AA M}^{(0)}\frac{\left(\gamma_{M}^{(1)}-2\gamma_{A}^{(1)}\right)^2}{2}
+\frac{4}{3}\xi_{N}^{(0)}g_{AA N}^{(0)}\frac{\left(\gamma_{N}^{(1)}-2\gamma_{A}^{(1)}\right)^2}{2}
\, ,\\  \nonumber
\frac{1}{640 \pi ^2}&=\xi_{\Sigma}^{(0)}g_{AA \Sigma}^{(0)}\frac{\left(\gamma_{\Sigma}^{(1)}-2\gamma_{A}^{(1)}\right)^2}{2}+
\xi_{K}^{(0)}g_{AA K}^{(0)}\frac{\left(\gamma_{K}^{(1)}-2\gamma_{A}^{(1)}\right)^2}{2}\\
&-\frac{2}{3}\xi_{M}^{(0)}g_{AA M}^{(0)}\frac{\left(\gamma_{M}^{(1)}-2\gamma_{A}^{(1)}\right)^2}{2}
-\frac{2}{3}\xi_{N}^{(0)}g_{AA N}^{(0)}\frac{\left(\gamma_{N}^{(1)}-2\gamma_{A}^{(1)}\right)^2}{2}
\,.
\label{eq_OPEcheck_1loop_x2log2}
\end{align}
where $\Sigma$ and $K$,  $M_i$ and $N_i$ are the operators with engineering dimension four that contributed to~\eqref{eq_OPEcheck_tree_x2log1} and~\eqref{eq_OPEcheck_tree_x2log2} at tree-level.  Also~\eqref{eq_OPEcheck_1loop_x2log1} and~\eqref{eq_OPEcheck_1loop_x2log2} are in agreement with the results~\eqref{eq_app_Delta4_singlet},~\eqref{eq_app_Delta4_doublet},~\eqref{eq_app_OPE} and \eqref{eq_app_vev_dim4} in the appendix.

\section{Outlook}

\label{sec_outlook}

In CFTs with moduli spaces, the operator spectrum and OPE coefficients are not independent of the particle spectrum and interactions on the moduli space. At least a subset of the relations between these quantities is obtained by expressing correlation functions in the moduli space in terms of an infinite number of one-point functions by repeated application of the OPE. 

In this paper we have studied the simplest set of such constraints, those that arise when considering a two-point function of local operators by comparing the OPE channel with the exchange of the asymptotic states in the broken vacuum, equation~(\ref{bootstrap}). Our most conceptual point concerns the convergence of the OPE, which we found admits an infinite radius of convergence under mild assumptions. We also studied a concrete example, the real $ABC$ model in $4-\epsilon$ dimensions, to one-loop order. We found that already at low orders in perturbation theory, \eqref{bootstrap} enforces nontrivial identities between the CFT data of the theory to arbitrarily high order in the coupling. We provided a systematic framework to write such relations order by order in perturbation theory  in arbitrary weakly coupled models, and verified several of the constraints we found explicitly in the example at hand.

Overall, our analysis provides an abstract viewpoint, not based on a Lagrangian description, on why CFTs with moduli space are nongeneric in the space of theories. Yet, we did not find a systematic approach to study the  bootstrap constraints in full generality and at the nonperturbative level. We still seem far from answering the longstanding question of which conditions must the CFT data of a theory obey for it to admit a moduli space. Below we list some possible future directions.

The most obvious next step is perhaps to extend our analysis of the bootstrap equation~\eqref{bootstrap} to a more constrained example, such as planar $\mathcal{N}=4$ Super-Yang-Mills theory. This promises to be a rich problem due to the possibility of exploring both the weak and strong coupling regime, using supergravity as in \cite{Bianchi:2001de}, and because of the potential interplay with integrability. (While this work was under completion, \cite{Ivanovskiy:2024vel} appeared where an analysis of this setup was initiated).

Besides studying additional explicit examples, it would be desirable to make the abstract bootstrap problem more systematic. A key technical obstacle to this aim is that the OPE channel of~\eqref{bootstrap} does not admit any positivity property; in fact we argued that the OPE expansion must necessarily yield a nontrivial oscillating series to be consistent with a regular long distance limit. Below we thus discuss some potential alternative strategies to decode the properties of the data of a CFT with a moduli space.

The simplest alternative is probably to consider the S-matrix in the broken vacuum, which obeys well known unitarity and crossing constraints, as in the S-matrix bootstrap literature (see~e.g.~\cite{Kruczenski:2022lot} for a recent review).
The existence of a moduli space would then be implemented by demanding the existence of a massless dilaton, whose low energy interactions are constrained by the nonlinear realization of the symmetry~\cite{Salam:1969bwb,Isham:1970gz}. See~\cite{Karateev:2022jdb} for progress on a related problem. Additional constraints might arise including form-factors of the stress tensor in the bootstrap system as in~\cite{Karateev:2019ymz}, and imposing tracelessness.

In a companion paper~\cite{Cuomo:2024fuy}, we take a very different approach. We take as an input the \emph{effective field theory} (EFT) for the dilaton and the other massless moduli \cite{Salam:1969bwb,Isham:1970gz,Gretsch:2013ooa}, and, rather than expanding around the broken vacua, we look for backgrounds that correspond to primary states in radial quantization. This approach has the advantage that the tuning required for the existence of a moduli space, the absence of a potential of the dilaton, is an explicit input in the construction of the EFT. Generalizing the ideas of \cite{Hellerman:2017veg}, we establish in full generality that, when an internal charge $Q$ is spontaneously broken in the moduli space, the lowest dimensional operators charged under the latter have scaling dimension growing linearly with the charge $\Delta_{min}(Q)\propto Q$ for $Q\gg 1$. The result is in general unrelated to the existence of BPS operators, and it applies also when $Q$ is not an $R$-charge, as we verify in several examples. This behaviour is to be contrasted with the scaling $\Delta_{min}(Q)\propto Q^{\frac{d}{d-1}}$ which is expected to hold in generic CFTs~\cite{Hellerman:2015nra,Monin:2016jmo}, and thus is a nontrivial reflection of the existence of a moduli space. In \cite{Cuomo:2024fuy} we will also discuss additional aspects of the connection between large charge operators and moduli spaces (see also \cite{Jafferis:2017zna,Caetano:2023zwe}), and list some interesting directions for generalizing this approach.

We also mention that in \cite{Caron-Huot:2014gia} it was also argued that, in $\mathcal{N}=4$ SYM, the massive spectrum of the Coulomb branch is related to the cusp anomalous dimensions of supersymmetric Wilson lines. It would be interesting to understand if there exist other quantitative connections between conformal defects and the S-matrix in the broken vacuum in conformal gauge theories.

Finally, we cannot resist but mention that it would be interesting if any of these approaches could shed light on the longstanding question of the existence of a dilaton in the conformal window of QCD \cite{Ellis:1970yd,Appelquist:2010gy} (see \cite{LSD:2023uzj} and ref.s therein for a summary of lattice searches). A priori, as emphasized in \cite{Gorbenko:2018ncu,Zan:2019idm}, the existence of a moduli space in a non-supersymmetric theory would be a rather finely tuned phenomenon. In particular, it is now understood that the spontaneous breaking of conformal symmetry is not related to the fixed point merger mechanism, which is believed to take place at the lower edge of the conformal window, as the examples studied in \cite{Pomoni:2009joh,Benini:2019dfy} show (see however \cite{Faedo:2024zib} for some holographic examples of dilatons at fixed-point merger points). Yet, there exist some examples of moduli spaces in large $N$ theories \cite{Rabinovici:1987tf,PhysRevLett.52.1188,Chai:2020zgq,Chai:2020onq}, and thus it is not logically impossible that a flat direction accidentally opens up in the large $N_c$ limit of QCD at some value of $N_f/N_c$. An elegant speculative scenario would be that such a flat direction emerges precisely at the lower edge of the conformal window, so that the conformal and spontaneously broken phases would be continuously connected.

.

\section*{Acknowledgments}
We thank Zohar Komargodski and  Yifan Wang for useful discussions.  The work of GC was supported by the Simons Foundation grant 994296 (Simons Collaboration on Confinement and QCD Strings). 
The work of LR was supported in part by the National Science Foundation under Grant NSF PHY-2210533 and by the  Simons Foundation under Grant 681267 (Simons Investigator Award).

\appendix

\section{Technical details of the perturbative calculation }

\label{app_ABC_CFT_data}

\subsection{Some details on the scheme}

We work in bare perturbation theory with the Lagrangian
\begin{equation}\label{eq_ABC_action_bare}
\begin{split}
\mL &=\frac{1}{2}(\pd a)^2+\frac{1}{2}(\pd b)^2+\frac{1}{2}(\pd c)^2+
\bar{\tilde{a}}_i\slashed{\pd}\tilde{a}_i+\bar{\tilde{b}}_i\slashed{\pd}\tilde{b}_i+
\bar{\tilde{c}}_i\slashed{\pd}\tilde{c}_i
\\ &
+\frac{g_0}{2}\left[a\left(\bar{\tilde{b}}_i\tilde{c}_i+\bar{\tilde{c}}_i\tilde{b}_i\right)
+b\left(\bar{\tilde{c}}_i\tilde{a}_i+\bar{\tilde{a}}_i\tilde{c}_i\right)
+c\left(\bar{\tilde{a}}_i\tilde{b}_i+\bar{\tilde{b}}_i\tilde{c}_i\right)\right]
\\&
+\frac{g^2_0}{8}\left(a^2b^2+b^2 c^2+c^2a^2\right)\,,
\end{split}
\end{equation}
where $g_0$ is the bare coupling and in this section only we work in terms of bare fields. All calculations are done for $i=1,...,N_\Psi$ where $N_\Psi=1/4$.

The relation between the bare coupling and the renormalized (physical) one is expressed through an ascending series of poles at $\epsilon=0$:
\begin{equation}
g_0=\mu^{\epsilon/2}\left[g+\frac{\delta g}{\epsilon}+\frac{\delta_2 g}{\epsilon^2}+\ldots\right]\,, 
\end{equation}
where $\mu$ is the sliding scale. To two-loop order, we find
\begin{align}
\delta g&=\frac{5 g^3}{4(4\pi)^2}-\frac{9 g^5}{16(4\pi)^4}+O\left(\frac{g^7}{(4\pi)^6}\right)\,, \\
\delta_2 g&=\frac{75 g^5}{32(4\pi)^4}+O\left(\frac{g^7}{(4\pi)^6}\right) \,,
\end{align}
Requiring that the bare coupling is independent of $\mu$, we extract the beta-function~\eqref{eq_ABC_beta_g}.

In the minimal subtraction scheme, the wave-function renormalization of a composite or fundamental operator $\mO$ is written as a series of ascending poles with no finite term:
\begin{equation}
    Z_{\mO}=1+\frac{z_{\mO}^{(1)}(g)}{\epsilon}
   +\frac{z_{\mO}^{(2)}(g)}{\epsilon^2}+\ldots\,,
\end{equation}
where each coefficient is given in a perturbative series as $z_{\mO}^{(k)}(g)=\#g^{2k}+\# g^{2k+2}+\ldots$. The wave-function is related to the anomalous dimensions by
\begin{equation}
\Delta_{\mO}-\Delta_{\mO}\vert_{classical}=Z_{\mO}^{-1}\frac{\pd Z_{\mO}}{\pd g}\left(-\frac{\epsilon}{2}g+\beta_g\vert_{d=4}\right)\,,
\end{equation}
where $\Delta_{\mO}\vert_{classical}$ is the engineering dimension of the operator.  We then define the renormalized operator as
\begin{equation}\label{eq_app_ABC_wavefunction}
\mO=\mu^{\Delta_{\mO}\vert_{classical}-\Delta_{\mO}}Z_{\mO}\mO_{ren.}\,,
\end{equation}
where  the subscript distinguishes the renormalized operator from the bare one and the factor of the sliding scale upfront is chosen so that $\mO_{ren}$ has dimension $\Delta_{\mO}$.  This ensures that correlation functions of renormalized operators at the fixed point are $\mu$-independent. We provide explicit results for the wave-functions and scaling dimensions of several operators in the next section.

To compute correlation functions in the moduli space, we simply need to shift $a\rightarrow a+v_0$ in the Lagrangian~\eqref{eq_ABC_action_bare}, where $v_0$ is the expectation value of the bare field:
\begin{equation}\label{eq_ABC_action_bare2}
\begin{split}
\mL&=\frac{1}{2}(\pd a)^2+\frac{1}{2}\sum_{\pm}\left[(\pd x_\pm)^2+m_0^2 x_{\pm}^2\right]+\bar{\tilde{a}}\slashed{\pd}\tilde{a}
+\sum_{\pm}\left[\bar{\tilde{x}}_{\pm}\left(\slashed{\pd}\mp m_0\right)\tilde{x}_{\pm}\right]\\&
-\frac{g_0}{2}a\left(\bar{\tilde{x}}_{-}\tilde{x}_{-}-\bar{\tilde{x}}_{+}\tilde{x}_{+}\right)-
\frac{g_0}{2}x_-\left(\bar{\tilde{x}}_{-}\tilde{a}+\bar{\tilde{a}}\tilde{x}_{-}\right)
+\frac{g_0}{2}x_+\left(\bar{\tilde{x}}_{+}\tilde{a}+\bar{\tilde{a}}\tilde{x}_{+}\right)
\\&
+\frac{g_0 m_0}{2}a(x_+^2+x_-^2)+\frac{g^2_0}{8}a^2(x_+^2+x_-^2)+\frac{g^2_0}{32}(x_+^2-x_-^2)^2\,,
\end{split}
\end{equation}
where $m_0=g_0 v_0/2$, we suppressed the fermion flavor index $i$ and $X_{\pm}$ are defined in~\eqref{eq_ABC_X_def}.

Because of~\eqref{eq_app_ABC_wavefunction} the vev of the bare operator $A$ is related to the vev of the renormalized field~\eqref{eq_vev_a} by
\begin{equation}\label{eq_vev_bare_renormalized}
    \langle a_{ren.}\rangle= \mu^{\Delta_A-\frac{d-2}{2}}Z_A^{-1}\langle a\rangle \,.
\end{equation}
Equivalently, the bare vev is related to the moduli space coordinate $v$ as
\begin{equation}
Z_{A}^{-1}v_0=v^{\frac{d-2}{2}}\left(\frac{m}{\mu}\right)^{\Delta_A-\frac{d-2}{2}}\,,
\end{equation}
where $Z_A$ and $\Delta_A$ are, respectively, the wave-function renormalization and the scaling dimension of the field, given in~\eqref{eq_app_Za} and~\eqref{eq_app_DeltaA} of the the next subsection. 

In the following subsections, as well as in the main text, we will never consider bare fields again. We therefore remove the subscript $ren.$ and indicate renormalized operators with the same symbols as their bare counterpart.

\subsection{Scaling dimensions}

We obtained the anomalous dimensions and wave-functions at the fixed-point \eqref{eq_ABC_g_fix} for some operators with engineering dimension $1$, $2$ and $4$ (in $d=4$) using the formulas in the appendix of \cite{Fei:2016sgs}.\footnote{To compute the anomalous dimension of operators of engineering dimension $4$, we included in the action all the dimension $4$ scalar operators invariant under the sign flip of two arbitrary superfields. We then obtained the beta functions of the corresponding couplings using the formulas in~\cite{Fei:2016sgs}, and computed the one-loop dilation operator restricted to this subsector from the derivatives of the beta function with respect to the coupling constants.} We list the results below.

Operators are organized in irreducible representations of the discrete group. The fundamental fundamental fields $(A,B,C)$ are grouped into an irreducible three-dimensional representation. We find that the corresponding wave-function renormalization, to two-loop order, is given by
\begin{align}\label{eq_app_Za}
Z_A&=1-\frac{g^2}{4(4\pi)^2\epsilon}-\frac{g^4}{(4\pi)^4}\left(
\frac{9}{32\epsilon^2}-\frac{1}{16\epsilon}\right)+O\left(\frac{g^6}{(4\pi)^6}\right)\,.
\end{align}
The scaling dimension follows from~\eqref{eq_app_ABC_wavefunction}:
\begin{equation}\label{eq_app_DeltaA}
\Delta_A=\frac{d-2}{2}+\frac{g_*^2}{4(4 \pi) ^2}-\frac{g_*^4}{8(4 \pi) ^2}+O\left(\frac{g_*^6}{(4 \pi) ^6}\right)=1-\frac{2\epsilon}{5}+\frac{2 \epsilon ^2}{125}+O\left(\epsilon^3\right)
\end{equation}

Let us now consider operators with classical dimension equal to $2$. There are two independent superconformal primary operators, which are neutral under the $\mathds{Z}_2$ group. These are conveniently grouped into a singlet and a doublet irrep.s of the permutation group $S_3$:
\begin{itemize}
\item The singlet is $S\equiv A^2+B^2+C^2$,
with wave-function renormalization
\begin{align}
Z_S&=1-\frac{3g^2}{2(4\pi)^2\epsilon}-\frac{g^4}{(4\pi)^4}\left(
\frac{3}{4\epsilon^2}-\frac{1}{\epsilon}\right)+O\left(\frac{g^6}{(4\pi)^6}\right)\,.
\end{align}
The corresponding scaling dimension is
\begin{equation}\label{eq_app_DeltaS}
\Delta_S=d-2+
\frac{3g^2_* }{2(4\pi)^2}-2\frac{g^4_* }{(4\pi)^4}+O\left(\frac{g_*^6}{(4 \pi) ^6}\right)=2-\frac{2 \epsilon }{5}-\frac{13 \epsilon ^2}{125}+O\left(\epsilon^3\right)\,;
\end{equation}
\item The doublet can be written as $\{D_1,\,D_2\}\equiv\{A^2-B^2,\,(A^2+B^2-2C^2)/\sqrt{3}\}$, where we chose a basis such that the two-point function is proportional to the identity $\langle D_i D_j\rangle\propto\delta_{ij}$; its wave-function reads
\begin{align}
Z_D&=1+0-\frac{g^4}{8(4\pi)^4\epsilon}+O\left(\frac{g^6}{(4\pi)^6}\right)\,.
\end{align}
The scaling dimension follows
\begin{equation}\label{eq_app_DeltaD}
\Delta_D=d-2+0+\frac{g^4_* }{4(4\pi)^4}+O\left(\frac{g_*^6}{(4 \pi) ^6}\right)=2-\epsilon+\frac{\epsilon ^2}{25}+O\left(\epsilon^3\right)\,.
\end{equation}
\end{itemize}
Note also that the triplet $\{AB,\,BC,\,CA\}$  is a superconformal descendant of the fundamental field because of the equations of motion, that imply $\mathcal{Q}^2 A\sim BC$, etc. .  We checked that its dimension is $\Delta_A+1$ to two-loop order as expected. 

In the main text we will also consider some operators with engineering dimension $4$. The ones we need are neutral under the $\mathds{Z}_2$ action, and consist of two singlets and two doublets of $S_3$:
\begin{itemize}
\item the two singlets that diagonalize the one-loop dimension matrix are
\begin{align}\nonumber
\Sigma &=A^4+B^4+C^4-(A^2 B^2+B^2C^2+C^2 A^2)+O\left(\sqrt{\epsilon}\right)\,,\\
K&=A^4+B^4+C^4+6(A^2 B^2+B^2C^2+C^2 A^2)+O\left(\sqrt{\epsilon}\right)\,,
\end{align}
where the $O(\sqrt{\epsilon})$ corrections include terms with fermions and derivatives, that ensure that these operators do not overlap with descendants of $S$ or other primaries; to one-loop order, the scaling dimensions read
\begin{align}\nonumber
\Delta_{\Sigma}&=2(d-2)+0+O\left(\frac{g^4_*}{(4\pi)^4}\right)=4-2\epsilon+O\left(\epsilon^2\right)\,,\\
\Delta_{K}&=2(d-2)+\frac{7g^2_*}{(4\pi)^2}+O\left(\frac{g^4_*}{(4\pi)^4}\right)=4+\frac{4}{5}\epsilon+O\left(\epsilon^2\right)\,;
\label{eq_app_Delta4_singlet}
\end{align}
\item the two doublets are permutations of
\begin{align}\nonumber
M_1&=A^4-B^4-\frac{12}{7+\sqrt{73}}(A^2-B^2)C^2+O\left(\sqrt{\epsilon}\right)\,,\\
N_1&=
A^4-B^4-\frac{12}{7-\sqrt{73}}(A^2-B^2)C^2+O\left(\sqrt{\epsilon}\right)\,,
\end{align}
such that, working again in a basis where $\langle M_i M_j\rangle\propto\delta_{ij}$, the second component reads $M_2=(M_1\vert_{B\leftrightarrow C}-M_1\vert_{A\leftrightarrow C})/\sqrt{3}$ and similarly for $N_2$; the corresponding one-loop scaling dimensions read
\begin{align}\nonumber
\Delta_{M}=2(d-2)+\frac{\left(11-\sqrt{73}\right) g^2_*}{4 (4\pi )^2}+O\left(\frac{g^4_*}{(4\pi)^4}\right)=4-\frac{9+\sqrt{73}}{10}\epsilon+O\left(\epsilon^2\right)\,,\\
\Delta_{N}=2(d-2)+\frac{\left(11+\sqrt{73}\right) g^2_*}{4 (4\pi )^2}+O\left(\frac{g^4_*}{(4\pi)^4}\right)=4-\frac{9-\sqrt{73}}{10}\epsilon+O\left(\epsilon^2\right)\,.
\label{eq_app_Delta4_doublet}
\end{align}
\end{itemize}

Finally, to compute OPE coefficients in the next section we will also need the following three conformal primaries
\begin{align}\nonumber
    \widetilde{W}&=\frac{1}{2}\left[a\left(\bar{\tilde{b}}\tilde{c}+\bar{\tilde{c}}\tilde{b}\right)+b\left(\bar{\tilde{c}}\tilde{a}+\bar{\tilde{a}}\tilde{c}\right)+c\left(\bar{\tilde{a}}\tilde{b}+\bar{\tilde{b}}\tilde{c}\right)\right]+\frac{g}{4}\left(a^2b^2+b^2c^2+c^2a^2\right)+O\left(\epsilon\right)\,,\\
    \nonumber \tilde{T}_1&=\left[a\left(\bar{\tilde{b}}\tilde{c}+\bar{\tilde{c}}\tilde{b}\right)-b\left(\bar{\tilde{c}}\tilde{a}+\bar{\tilde{a}}\tilde{c}\right)\right]+g\left(a^2-b^2\right)c^2
    +O\left(\epsilon\right)\,,
    \\
    \label{eq_app_Wt}
    \tilde{T}_2&=
    \frac{1}{\sqrt{3}}\left(\tilde{T}_1\vert_{B\leftrightarrow C}-\tilde{T}_1\vert_{A\leftrightarrow C}\right)
    +O\left(\sqrt{\epsilon}\right)\,,
\end{align}
where $\{\tilde{T}_1,\tilde{T}_2\}$ form a doublet and $\widetilde{W}$ is a superconformal descendant of the superpotential $W$ at the fixed point. In~\eqref{eq_app_Wt} we neglected the fermion flavor index. For completeness, we report the corresponding one-loop scaling dimensions
\begin{align}\nonumber
    \Delta_{\widetilde{W}}&=4+O\left(\epsilon^2\right)\,,\\
    \Delta_{\tilde{T}}&=4+\frac{7}{10}\epsilon+O\left(\epsilon^2\right)\,.
\end{align}
 The operators $\Sigma$, $K$, $M_i$, $N_i$, $\widetilde{W}$, $\tilde{T}_i$ are all the operators with engineering dimension $4$ that are invariant when the sign of two different superfields is reversed.

\subsection{OPE coefficients}

Let us now present the OPE of the fundamental field to order $O\left(\epsilon\right)$. For convenience we define a vector $(V_1,V_2,V_3)=(A,B,C)$, and an invariant tensor $t_{ab,i}$ where $a,b=1,2, 3$ and $i=1,2$ are, respectively, triplet and doublet indices of the $S_4$ symmetry group. Explicitly, only the components of $t_{ab,i}$ which are diagonal in the first two indices are non-zero, and read:
\begin{equation}
\left(\begin{array}{cc}
  t_{11,1}   & t_{11,2}  \\
  t_{22,1}   &  t_{22,2}\\
  t_{33,1} & t_{33,2}
\end{array}\right)=
\left(\begin{array}{cc}
  1   & \frac{1}{\sqrt{3}}  \\
  -1   &  \frac{1}{\sqrt{3}}\\
  0 & -\frac{2}{\sqrt{3}}
\end{array}\right)\,.
\end{equation}
Given this definition, we can compactly write the contribution to the OPE from the operators listed in the previous section:
\begin{equation}\label{eq_app_OPE}
\begin{split}
V_a(x)V_b(0) &\supset g_{AAS}|x|^{\Delta_S-2\Delta_A}\delta_{ab}S(0) +g_{AAD}|x|^{\Delta_S-2\Delta_A}t_{ab,i}D_i(0) \\
&+g_{AA\Sigma}|x|^{\Delta_\Sigma-2\Delta_A}\delta_{ab}\Sigma(0)
+g_{AAK}|x|^{\Delta_{K}-2\Delta_A}\delta_{ab}K(0)\\
&+g_{AA M}|x|^{\Delta_{M}-2\Delta_A}t_{ab,i} M_i(0)
+g_{AA N}|x|^{\Delta_{N}-2\Delta_A}t_{ab,i} N_i(0)\\
&+g_{AA\widetilde{W}}|x|^{\Delta_{\widetilde{W}}-2\Delta_A}\delta_{ab}\widetilde{W}(0)
+g_{AA \tilde{T}}|x|^{\Delta_{\tilde{T}}-2\Delta_A}t_{ab,i} \tilde{T}_i(0)
\,.
\end{split}
\end{equation}
Below we compute the explicit value of the OPE coefficients in~\eqref{eq_app_OPE}. 

In general, to compute the OPE coefficients we need to know the two-point function of the composite operator $\mO$ of interest, from which we extract the normalization factors that relates canonically normalized operators and those in the minimal subtraction scheme, and the three-point function $\langle A A \mO\rangle$. Explicitly, we define the two-point function normalization of the lowest components of the superfields in the minimal subtraction scheme as
\begin{equation}
\qquad\langle \mO(x) \mO(0)\rangle=\frac{\mathcal{N}_{\mO\mO}}{|x|^{2\Delta_{\mO}}} \,.
\end{equation}
The OPE coefficients of $A$ and an operator $\mathcal{O}$ in then obtained from the three-point function
\begin{align}\label{eq_app_3pt_form}
    \langle A(x)A(y)\mathcal{O}(0)\rangle=
    \frac{\mathcal{N}_{\mO\mO}g_{AA\mO}}{|x-y|^{2\Delta_A-\Delta_{\mO}}|y|^{\Delta_{\mO}}|x|^{\Delta_{\mO}}}\,.
\end{align}
Note that in our scheme we do not need to know the normalization of the two-point function of $A$. In practice, as usual, the expansion of the anomalous dimension will lead to logarithms of the distance between the operators in perturbation theory .

The calculation of the OPE coefficients of the operators with classical dimension $2$, i.e. $S$ and $D_i$, is straightforward, since they have different quantum numbers under the internal symmetry and thus we do not have to worry about operator mixing. At tree-level, the result immediately follows from the free theory OPE $A\times A\supset A^2$ and the identity
\begin{eqnarray}\label{eq_app_AA_dec}
    A^2=\frac13S+\frac12 D_1+\frac{1}{2\sqrt{3}}D_2\,.
\end{eqnarray}
At one-loop, one finds
\begin{align}\label{eq_app_N_SS}
\mathcal{N}_{SS}& = \frac{3}{8 \pi ^4}+\frac{3 \epsilon  (\gamma +\log \pi -2)}{20 \pi ^4}+O\left(\epsilon^2\right)\,,\\
\mathcal{N}_{DD} &=  \frac{1}{4 \pi ^2}+\frac{\epsilon  (5 \gamma +5 \log \pi -1)}{20 \pi ^2}+O\left(\epsilon^2\right)\,.
\label{eq_app_N_DD}
\end{align}
The calculation of the tree-point functions is straightforward and one obtains:
\begin{align}
g_{AAS}&= \frac13+\frac{1}{15} \epsilon  (\gamma +\log \pi )+O\left(\epsilon ^2\right)\,,\\ 
g_{AAD}& =
\frac{1}{2}-\frac{1}{20} \epsilon  (\gamma +\log \pi )
+O\left(\epsilon ^2\right)\,.
\end{align}

To efficiently compute the OPE coefficients of the operators with engineering dimension $4$, we note that to order $O(g^2)$ the only composite operators made of more than two fundamental fields whose three-point three-point function with two $a$'s is nonvanishing are $a^2b^2$ and $a\left(\bar{\tilde{b}}\tilde{c}+\bar{\tilde{c}}\tilde{b}\right)$. Since we work only at the leading nontrivial order, we may compute the OPE coefficients of these two operators directly, and then decompose them in terms of the operators that diagonalize the dilation operator listed in the previous section, similarly to~\eqref{eq_app_AA_dec}. There is however as small subtlety: we should make sure that the operators that we use have zero overlap with lower dimensional primaries such as $a^2$ and $b^2$. To the order of interest, this is achieved considering the combinations
\begin{align}\label{eq_app_O3}
\mO_{3}&=a\left(\bar{\tilde{b}}\tilde{c}+\bar{\tilde{c}}\tilde{b}\right)-\frac{g}{(4 \pi )^2}(\partial a)^2\,,\\
    \mO_{4}&=a^2b^2-\frac{g^2}{(4 \pi )^4}(\partial a)^2-\frac{g^2}{(4 \pi )^4}(\partial b)^2\,,
    \label{eq_app_O4}
\end{align}
The subtracted terms ensure that three-point functions take the expected structure~\eqref{eq_app_3pt_form}.\footnote{In practice, we can neglect the additional terms in~\eqref{eq_app_O3} and~\eqref{eq_app_O4} if we content ourselves with computing the three-point function~\eqref{eq_app_3pt_form} to leading order at small $|x-y|$, which is enough to extract the OPE coefficient.} 
Using the results for the tree-level two-point functions,
\begin{equation}
    \langle \mO_3(x)\mO_3(0)\rangle=\frac{1}{8\pi^6|x|^4}+O\left(\epsilon\right)\,,\qquad
    \langle \mO_4(x)\mO_4(0)\rangle=\frac{1}{64\pi^8|x|^4}+O\left(\epsilon\right)\,,
\end{equation}
and computing the three-point functions to $O(g^2)$, we obtain the contribution of $\mO_3$ and $\mO_4$ to the OPE of the fundamental field
\begin{eqnarray}
    A(x)A(0)\supset \left[\frac{g}{16}+O\left(g^3\right)\right]x^2\mO_3(0)+\left[\frac{g^2}{32}+O\left(g^4\right)\right]x^2\mO_4(0)
\end{eqnarray}
To obtain the result for the OPE coefficients in~\eqref{eq_app_OPE}, we simply need to decompose $\mO_3$ and $\mO_4$ in terms of a complete basis of operators of classical dimension $4$ that can appear in the OPE, i.e. $\Sigma$, $K$, $M_i$, $N_i$, $\widetilde{W}$ and $\tilde{T}_i$. We find
\begin{align}
g_{AA\Sigma}& =-g_{AAK}=-\frac{\pi^2}{105}  \epsilon +O\left(\epsilon ^2\right)\,,\\ 
g_{AAM}&=-g_{AAN}=\frac{\pi ^2 \epsilon }{10 \sqrt{73}}+O\left(\epsilon ^2\right)\,,\\
g_{AA\widetilde{W}}&=\frac{\pi  \sqrt{\epsilon }}{3 \sqrt{10}}\left[1+O\left(\epsilon\right)\right]\,,\\
g_{AA\tilde{T}}&=\frac{\pi  \sqrt{\epsilon }}{4 \sqrt{10}}\left[1+O\left(\epsilon\right)\right]\,.
\end{align}
Note that the subtractions in~\eqref{eq_app_O3} and~\eqref{eq_app_O4} can be neglected to this order when rewriting $\mO_3$ and $\mO_4$ in terms of operators that diagonalize the dilation operator.

Finally, for completeness we provide the expression for the OPEs of canonically normalized operators for the operators in the first three lines of~\eqref{eq_app_OPE}. To this aim, we need the normalization of the two-point function of the fundamental field
\begin{equation}
\mathcal{N}_{AA} =  \frac{1}{4 \pi ^2}+\frac{\epsilon  (2 \gamma +2 \log \pi -1)}{20 \pi ^2}+O\left(\epsilon^2\right)\,, 
\end{equation}
the two point functions~\eqref{eq_app_N_SS} and~\eqref{eq_app_N_DD} of $S$ and $D_i$, and the analogous results for $\Sigma$, $K$, $M_i$ and $N_i$:
\begin{align}
&\mathcal{N}_{\Sigma\Sigma}=\frac{21}{64 \pi ^8}+O\left(\epsilon\right)\,,&&
\mathcal{N}_{KK}=\frac{63}{32 \pi ^8}+O\left(\epsilon\right)\,,\\
&\mathcal{N}_{MM}=\frac{73-7 \sqrt{73}}{64 \pi ^8}+O\left(\epsilon\right)\,,&&
\mathcal{N}_{NN}=\frac{7 \sqrt{73}+73}{64 \pi ^8}+O\left(\epsilon\right)\,.
\end{align}
The OPE coefficients for canonically normalized operators are then obtained via $\lambda_{AA\mO}=g_{AA\mO}\sqrt{\mathcal{N}_{\mO\mO}}/\mathcal{N}_{AA}$. We find
\begin{align}
&\lambda_{AAS} =\sqrt{\frac{2}{3}}-\frac{1}{5} \sqrt{\frac{2}{3}} \epsilon+O\left(\epsilon ^2\right)\,,
&&\lambda_{AAD} =1+\frac{\epsilon }{10}
+O\left(\epsilon ^2\right)\,,\\ 
&\lambda_{AA\Sigma}=-\frac{\epsilon }{10 \sqrt{21}} +O\left(\epsilon ^2\right)\,,
&&\lambda_{AAK}=\frac{\epsilon }{5 \sqrt{14}} +O\left(\epsilon ^2\right)\,,\\ 
&\lambda_{AAM}=\frac{\sqrt{6+\frac{42}{\sqrt{73}}} \,\epsilon }{10 \left(\sqrt{73}+7\right)}+O\left(\epsilon ^2\right)\,,
& &\lambda_{AAN}=-\frac{\sqrt{6-\frac{42}{\sqrt{73}}}\, \epsilon }{10 \left(\sqrt{73}-7\right)}+O\left(\epsilon ^2\right)\,.
\end{align}

\subsection{One-point functions}

In the minimal subtraction scheme, the one-point functions of the two lowest dimensional nontrivial operators to two loop order are given by
\begin{equation}\label{eq_app_1pt_pre}
\langle S\rangle=v^{d-2}m^{\Delta_S-(d-2)}
\hat{\xi}_S\,,\qquad
\langle D_1\rangle =\sqrt{3}\langle D_2\rangle=v^{d-2}m^{\Delta_D-(d-2)}\hat{\xi}_D\,,
\end{equation}
where
\begin{align}\nonumber
\hat{\xi}_S &=1+\frac{\epsilon}{5}\left[\gamma -1-\log (4 \pi )\right] +\frac{\epsilon^2}{3000}\left[60 \gamma ^2
-12 \gamma  (7+20 \log 2+10 \log \pi )
-25 \pi ^2
\right.
\\ \label{eq_app_1pt_S}
&\left. 
+54+168 \log 2+84 \log \pi +24 \log (1024) \log (2)+60 \log (\pi ) \log (16 \pi )
\right]+O\left(\epsilon^3\right)
\,,\\[0.5em] \nonumber
\hat{\xi}_D &=1+\frac{\epsilon}{10}  (1-\gamma +\log 4 \pi )+
\frac{\epsilon^2}{6000}\left[
30 \gamma ^2-12 \gamma  (23+\log 1024+5 \log \pi )+25 \pi ^2+396
\right.\\ \label{eq_app_1pt_D}
&\left.
+276 \log \pi +138 \log 16
+30 \log ^2(\pi )+30 \log (16) \log (\pi )+6 \log (16) \log (32)
\right]+O\left(\epsilon^3\right)
\,.
\end{align}

In the basis discussed in the former section, the tree-level one-point functions of the operators with engineering dimension $4$ are particularly simple and read
\begin{equation}\label{eq_app_vev_dim4}
\langle \Sigma\rangle=\langle K\rangle=\langle M_1\rangle=\langle N_1\rangle=\sqrt{3}\langle M_2\rangle=\sqrt{3}\langle N_2\rangle=\langle A^4\rangle= v^4\left[1+O\left(\epsilon\right)\right]\,.
\end{equation}
 
\subsection{Fourier transform of the one-loop two-point functions}\label{app_Fourier}

In this section we compute the Fourier transform of the two-point functions~\eqref{eq_aa2pt_1loop_mom} and~\eqref{eq_bb2pt_1loop_mom}. 

First, notice that to obtain the four-dimensional Fourier transform of a function $f=f(p^2)$ we can always perform the angular integration first and recast the integration as
\begin{equation}
\begin{split}
    \int\frac{d^4p}{(2\pi)^4}e^{-ip x}f(p^2) &=\frac{1}{4\pi^3}\int_{0}^{\infty} dp\,p^3 f(p^2)\int_0^{\pi} d\theta\sin^2\theta e^{-ip|x|\cos\theta} \\
    &=\frac{1}{(2\pi)^2|x|}\int_0^{\infty}dp\,p^2 f(p^2)J_1(p|x|)\,,
    \end{split}
\end{equation}
where $J_1$ is a Bessel function of the first kind. The result~\eqref{eq_xx_2pt_1loop} then follows simply performing the integration explicitly (e.g. using \texttt{Mathematica}).

Let us now discuss the Fourier transform of~\eqref{eq_aa2pt_1loop_mom}. This consists of two terms
\begin{equation}
\begin{split}
\frac{(4\pi)^2}{g^2}\langle a(x)a(0)\rangle^{1-loop}=\frac{1}{(2\pi)^2|x|}\int_0^{\infty}dp\, J_1(p|x|)&\left\{
\frac{\sqrt{4 m^2+p^2}}{2p} \tanh ^{-1}\left(\frac{p}{\sqrt{4 m^2+p^2}}\right) \right.\\[0.4em]
&\left.+\frac{1}{4}\left[
\log \left(\frac{m^2}{4 \pi  }\right)+\gamma -2\right]
\right\}\,.
\end{split}
\end{equation}
The terms in the second line can be integrated explicitly and we obtain
\begin{equation}\label{eq_app_aa_short_1}
\frac{(4\pi)^2}{g^2}\langle a(x)a(0)\rangle^{1-loop}= \frac{\log \left(\frac{m^2}{4 \pi  }\right)+\gamma -2}{16\pi^2|x|^2}+
\frac{1}{(2\pi)^2|x|^2}F(m|x|)\,,
\end{equation}
where 
\begin{equation}\label{eq_app_F}
    F(m |x|)=|x|\int_0^{\infty}dp\, J_1(p|x|)
\frac{\sqrt{4 m^2+p^2}}{2p} \tanh ^{-1}\left(\frac{p}{\sqrt{4 m^2+p^2}}\right)\,.
\end{equation}
We were not able to compute the integral~\eqref{eq_app_F} in closed form. We will therefore content ourselves with expressing the result as a series for small $m^2 x^2$. To this aim, note that the short distance expansion of the correlator in position space cannot be obtained directly from the large $p$ expansion in momentum space, performing the Fourier transform term by term. Indeed the large $p$ expansion of the result~\eqref{eq_aa2pt_1loop_mom} in momentum space produces terms of the form $1/p^{2n}$, which are not integrable for $n\geq 2$. To proceed instead we apply the method of matched asymptotic expansions. Namely, we separate the integration into two regions $(0,a/|x|)$ and $(a/|x|,\infty)$ with $0<a\ll 1 $:
\begin{equation}
    F(m |x|)=F_1(m|x|,a)+F_2(m|x|,a)\,,
\end{equation}
where
\begin{align}
F_1(m|x|,a)&=|x|\int_0^{a/|x|} dp\, J_1(p|x|)
\frac{\sqrt{4 m^2+p^2}}{2p} \tanh ^{-1}\left(\frac{p}{\sqrt{4 m^2+p^2}}\right)\,,\\[0.4em]  
F_2(m|x|,a)&=|x|\int_{a/|x|}^\infty dp \, J_1(p|x|)
\frac{\sqrt{4 m^2+p^2}}{2p} \tanh ^{-1}\left(\frac{p}{\sqrt{4 m^2+p^2}}\right)\,.
\end{align}
Since for $p\in (0,a/|x|)$ we have $p|x|\ll 1$, the integral in the first region can be performed expanding the Bessel function in a power series and performing the integration term by term. Expanding the result for $m |x|\ll a\ll 1$ we find a double series of the form
\begin{equation}\label{eq_app_F1}
\begin{split}
F_1(m|x|,a)&=(m|x|)^0\left\{\frac{1}{16} a^2 [2 \log a-2 \log (m|x|)-1] +O\left(a^4\right)\right\}\\
&+m^2x^2\left\{\frac{1}{8} \left[
2 \log^2(m|x|)-2 \log (m|x|)+1-4\log (m|x|) \log a \right.\right.
\\&
\qquad\qquad\left.+2 \log a+2 \log^2 a\right]
\left.+\frac{1}{32} a^2 \left[\log (m|x|)-\log a\right]+O\left(a^4\right)\right\}\\&+O\left(m^4|x|^4\right)\,.
\end{split}
\end{equation}
Similarly, in the region $p\in (a/|x|,\infty)$ we can expand the term multiplying the Bessel function in a series valid at $p^2\gg m^2 $. Performing the integral term by term and expanding for $m |x|\ll a\ll 1$ we find 
\begin{equation}\label{eq_app_F2}
\begin{split}
F_2(m|x|,a)&=(m|x|)^0\left\{
\frac{1}{2} \left[\log 2-\log (m|x|)-\gamma \right]-
\frac{1}{16} a^2 [2 \log a-2 \log (m|x|)-1] +O\left(a^4\right)\right\}\\
&+m^2x^2\left\{
\frac{1}{4} \left[2 \log (m|x|)\log a -\log ^2 a-\log a+(2 \gamma-\log 4-1)  \log (m|x|)\right.\right.\\
&\qquad\quad\left.\left.
+(\gamma -1)^2+\log ^2 2+(1-\gamma)  \log 4\right]-\frac{1}{32} a^2 \left[\log (m|x|)-\log a\right]+O\left(a^4\right)\right\}\\
&+O\left(m^4|x|^4\right)\,.
\end{split}
\end{equation}
Summing~\eqref{eq_app_F1} and~\eqref{eq_app_F2} we see that the dependence on $a$ cancels. From~\eqref{eq_app_aa_short_1} we then obtain the first few coefficients of the expansion in~\eqref{eq_aa_2pt_1loop}:
\begin{align}\nonumber
 \alpha_{-1}&=0\,,\quad &\alpha_{0}&=\frac{1}{16\pi^2}\,,\quad
 &\alpha_{1}&=\frac{1}{128\pi^2}\,,\\ 
  \beta_{-1}&=-\frac{1}{8\pi^2}\,,\quad &\beta_{0}&=\frac{ \log \frac{m}{2}+\gamma -1}{8 \pi ^2}\,,\quad
& \beta_{1}&=\frac{\log \frac{m}{2}+\gamma -1}{64 \pi ^2}\,,
\label{eq_app_alpha_beta_pre}
\end{align} 
\vspace*{-1.5em}
 \begin{align}\nonumber
 \gamma_{-1}&=-\frac{\gamma +2+\log (\pi )}{16 \pi ^2}\,,\\
 \gamma_{0}&=\frac{2 \left(\log \frac{m}{4}+2 \gamma -2\right) \log m+3+2 \left[\gamma ^2+\log ^2 2-2 \gamma  (1+\log 2)\right]+\log 16}{32 \pi ^2}\,,
\label{eq_app_gamma} 
 \\
 \gamma_1&=
 \frac{8 \left(\log \frac{m}{4}+2 \gamma -2\right) \log m+7+8 \left[\gamma ^2+\log ^2 2-2 \gamma  (1+\log 2)\right]+8 \log 4}{1024 \pi ^2}\,.
  \nonumber
\end{align}

In practice, we can also obtain the coefficients of the $x^{2k}\log^2|x|$ and $x^{2k}\log|x|$ terms in~\eqref{eq_aa_2pt_1loop} from the large momentum expansion of the result~\eqref{eq_aa2pt_1loop_mom}. To this aim we note that for any integer $k\geq 0$ we have
\begin{align}
&\int d^4x \log|x|\,x^{2k} e^{ipx}=\frac{b_k}{p^{4+2k}} \,,\\
& \int d^4x \log^2 |x|\, x^{2k} e^{ipx}=
\frac{a_k}{p^{4+2k}}\left[\log |p|+da_k\right]\,,
\end{align}
where
\begin{align}
a_k&=(-1)^k \pi ^2 2^{2 k+4} \Gamma (k+2)\,k! \,,\\
da_k&=\gamma -H_k-\frac{1}{2 (k+1)}-\log 2\,,\\
b_k&= (-1)^{k+1} \pi ^2 2^{2 k+3} \Gamma (k+2)\,k! \,,
\end{align}
and $H_k$ is the $k$th harmonic number. We then compare to the large momentum expansion of the result~\eqref{eq_aa2pt_1loop_mom}
\begin{equation}
\langle a(p)a(-p)\rangle^{(1-loop)}=\frac{g^2}{(4\pi)^2}\left[
\frac{\gamma -2-\log \left(4 \pi  \right)}{4 p^2}
+
\sum_{k=0}^\infty m^{2+2k}\frac{\tilde{\alpha}_k\log|p|+\tilde{\beta}_k}{p^{4+2k}}
\right]\,,
\end{equation}
where
\begin{align}
\tilde{\alpha}_k&=(-1)^k 2^{2k}
\frac{ \Gamma \left(k+\frac{1}{2}\right)}{\sqrt{\pi } \Gamma (k+2)}\,,\\
\tilde{\beta}_k&=(-1)^{k+1} 2^{2 k-1}  \frac{\Gamma \left(\frac{1}{2}+k\right) \left(H_{k-\frac{1}{2}}-H_{k+1}+\log 4\right)}{\sqrt{\pi } \Gamma (k+2)}-\tilde{\alpha}_k\log m\,.
\end{align}
We conclude that, for $k\geq 0$, the coefficients $\alpha_k$ and $\beta_k$  in~\eqref{eq_aa_2pt_1loop} are given by
\begin{equation}\label{eq_app_alpha_beta}
\alpha_k=\frac{\tilde{\alpha}_k}{a_k}\,,\qquad
\beta_k=\frac{\tilde{\beta}_k-da_k\tilde{\alpha}_k}{b_k}\qquad
k\geq 0\,.
\end{equation}
Eq.~\eqref{eq_app_alpha_beta} agrees with the former results~\eqref{eq_app_alpha_beta_pre} for $k=0,\,1$.

Note finally that the Fourier transform of $x^{2k}$ with $k$ integer vanishes for $p^2>0$. Thus we cannot extract the $\gamma_k$'s in~\eqref{eq_aa_2pt_1loop} commuting the integral with the large momentum expansion of the one-loop result, as we did for the coefficients of the logarithmic terms.

\bibliography{refs}
	\bibliographystyle{JHEP.bst}

\end{document}